\documentstyle[12pt,psfig,fleqn,aabib]{l-aa}

\def\eg{\mbox{e.g.}}
\def\etal{\mbox{et~al.}\ }
\def\apgt{\ {\raise-.5ex\hbox{$\buildrel>\over\sim$}}\ }
\def\aplt{\ {\raise-.5ex\hbox{$\buildrel<\over\sim$}}\ }
\def\P{Paczy\'{n}ski}

\def\TY{Tutukov \& Yungelson~}

\def\pz{Portegies~Zwart}
\def\pe{\mbox{$P_{orb}$ -- $e$}~}

\newcommand{\pyr}{\mbox {{\rm yr$^{-1}$}}}

\newcommand{\mh}{\mbox {km s$^{-1}$~Mpc$^{-1}$}}
\newcommand{\msun}{\mbox{${\rm M}_\odot$}}
\newcommand{\myr}{ \mbox{ ${\rm M}_\odot \,{\rm yr}^{-1}$}}
\newcommand{\lsun}{\mbox{${\rm L}_\odot$}}
\newcommand{\rsun}{\mbox{${\rm R}_\odot$}}
\newcommand{\kms}{\mbox{${\rm km~s}^{-1}$}}
\newcommand{\elce}{\mbox{$\eta_{\rm ce}\lambda$}}
\newcommand{\md}{\mbox {$\dot {M}$}}
\newcommand{\ace}{\mbox {$\alpha_{\rm ce}$}}
\newcommand{\SPH}{\mbox {smoothed-particle hydrodynamics}}
\newcommand{\RLOF}{\mbox {Roche-lobe overflow}}

\newcommand{\I}{I\ }
\newcommand{\II}{II\ }

\newcommand{\ecc}{\mbox{$_{\it e}$}}
\newcommand{\ms}{\mbox{${\it ms}$}}
\newcommand{\gs}{\mbox{${\it gs}$}}

\newcommand{\he}{\mbox{${\it he}$}}
\newcommand{\co}{\mbox{${\it co}$}}
\newcommand{\ns}{\mbox{${\it ns}$}}
\newcommand{\bh}{\mbox{${\it bh}$}}

\begin{document}
\thesaurus{08.02.1; 
	   08.14.1;
	   08.16.6;
	   10.19.2;}

\title{Formation and Evolution of Binary Neutron Stars}
\author{Simon~F.~Portegies~Zwart \inst{1}
\and
Lev R. Yungelson \inst{1,2}}
\offprints {S.~F.~Portegies~Zwart}
\institute{Astronomical Institute {\em Anton Pannekoek},
Kruislaan 403, NL-1098 SJ Amsterdam, The Netherlands; spz@astro.uva.nl
\and
Institute of Astronomy of the Russian Academy of Sciences, 48 
Pyatnitskaya Str., 109017 Moscow, Russia; lry@inasan.rssi.ru}

\date {Received; accepted: }

\maketitle
\markboth{Simon~Portegies~Zwart \& Lev~Yungelson: \,
	  Formation and Evolution of Neutron Star Binaries}{}

\begin{abstract}

The formation and evolution of binaries which contain two neutron
stars or a neutron star with a black hole are discussed in detail.
The evolution of the distributions in orbital period and eccentricity
for neutron star binaries are studied as a function of time.  In the
model which fits the observations of high mass binary pulsars
best the deposition of orbital energy into common envelopes has to be
very efficient and a kick velocity distribution has to contain a
significant contribution of low velocity kicks.  The estimated age of
the population has to be between several 100\,Myr and 1~Gyr.  The
birthrate of binary neutron stars is $\sim 3.4 \times 10^{-5}$\,\pyr\
(assuming 100\% binarity) and their merger rate is $\sim 2 \times
10^{-5}$\,\pyr.  The merger rate of neutron star binaries is
consistent with the estimated rate of $\gamma$-ray bursts, if the
latter are beamed into an opening angle of a few degrees.  We argue
that PSR~B2303+46 is possibly formed in a scenario in which the common
envelope is avoided while for the other three known high-mass binary
pulsars a common envelope is required to explain their orbital period.

\end{abstract}

\section{Introduction}

Peculiarly enough, the first binary pulsar to be discovered, PSR~
B1913+16 (Hulse \& Taylor 1975), turned out to contain two neutron stars one of
which is visible as a recycled pulsar and the other is not visible.
The majority of about thirty recycled pulsars discovered so far in the
Galactic disc are found to have white dwarf companions. Four of these
systems may have a neutron star companion instead,
as is suggested by their mass. 

An adequate model of the population of Galactic neutron star binaries
is highly desirable, as their mergers are currently considered as the
most viable sources of detectable bursts of gravitational wave
radiation (\cite{clae77}) and possibly 
$\gamma$-rays (\P\ 1986).  Estimates of the merger rates of the Galactic 
binaries
which contain two neutron stars based on observations are 
strongly biased by the uncertainties in the
pulsar lifetime, beaming factor and the fraction of unobserved low-luminosity
pulsars (\cite{cl95}; \cite{vdhl96}).
Therefore, it is useful to investigate the Galactic population of
neutron star binaries by simulating their formation and the subsequent
orbital decay due to gravitational wave radiation and  
to fit the orbital parameters to observations. 
The population of binary neutron stars is 
a part of the more general family of descendants of massive binaries. 
Hence, an adequate model has to comply with observations also in the 
stages preceding the formation of neutron stars (see Meurs \& van den
Heuvel 1989 for an example of this approach).

After the pioneering studies of Clark et al.\ 1979, 
Kornilov \& Lipunov (1984a,b) and \cite*{dc87} the 
formation of single and binary neutron stars and the merger rate of the 
latter are modelled by, among others Tutukov \&
Yungelson (1993a,b), Portegies Zwart \& Spreeuw (1996), Lipunov \etal
(1996, 1997), Bagot (1996).

Since the paper by
Flannery \& van den Heuvel (1975), a lot of effort was spent on 
reproduction of the parameters of PSR~B1913+16.  But surprisingly, the rest
of the family didn't attract much attention. The primary aim  of most 
studies was to constrain the orbital parameters of {\em immediate}
progenitors of high-mass binary pulsars (HMBP, hereafter) 
and kicks imparted to neutron 
stars (see \eg\ Wijers et al.\ 1992, 
Yamaoka \etal 1993; Fryer \& Kalogera 1997; Bagot 1997; Lipunov et al. 1997).
No attempts were made to restrict the models by simultaneous confrontation 
with  all observed systems and to understand how typical they are in 
the total inferred population of double neutron stars.
Though only four HMBP are definitely known\footnote{
We omit the globular
cluster object B2127+11C which could be formed by a dynamical encounter
rather than by binary evolution (Phinney \& Sigurdsson 1991) 
and B1820-11 which may be young or
have a non-neutron star as companion (\cite{pv91}).} 
a better understanding of their
formation is helpful, as it may provide additional constraints on
the probable rate of events accompanying their
merger.
                                             
In Sect. 2 we discuss the formation scenarios for HMBP.
(We use the terms ``neutron star'' as a generic term for the remnant
of a high mass star and ``pulsar'' for a neutron star which is
observable in radio.)
Particular attention is 
given to the revision of the ``standard'' scenario suggested by Brown (1995).
Section 3 gives the results of our population synthesis
computations and the inferred 
orbital-period--eccentricity diagram for binary 
neutron stars. This probability diagram 
provides a unique opportunity to compare the combined results of evolutionary
computations and population synthesis with observations of 
HMBP. A requirement for pulsars to be located in the regions 
of this diagram with high likelihood
also allows to constrain the progenitor population and its evolution.
We derive the estimate of the merger rate of the Galactic neutron stars 
from the model which fits best to the observed population.
Discussion follows in Sect.~4 and conclusions are summarized in Sect.~5. 

\section{Formation of binary neutron stars}
\subsection{Scenarios of formation}
Neutron stars are believed to be born in the supernova  
explosions triggered by the iron-core collapse in massive stars.
The range of progenitor masses of neutron stars in isolation assumed in 
the present paper is 8 to 40 \msun. However, the lower limit may be 
actually  
even higher  than 10 \msun; Ritossa \etal (1996) demonstrated  
that a 10 \msun\ star ends its evolution as 1.26 \msun\ oxygen-neon white 
dwarf. This would require a reduction of the model occurrence rate of 
supernovae from single stars by $\sim 30\%$.
If a star is stripped from its hydrogen envelope due to Roche-lobe overflow
in a close binary the lower mass limit for the
formation of a neutron star increases to about 12~\msun. 
The necessary condition for the formation of a neutron star is 
the requirement for the helium star remnant of a binary component to have a 
mass in excess
of about 2.2 to 2.5~\msun (Tutukov \& Yungelson 1973; Habets 1985), which 
enables the formation of a carbon-oxygen core of a Chandrasekhar mass. 
The upper limit is less well known. The main factors contributing 
to this are uncertain stellar mass-loss rates,
poor understanding of mixing processes in stellar interiors, the 
processes during the supernova itself and the uncertainty concerning the
equation of state of the compact object which results from the
supernova (see \eg\ Brown et al.\ 1996). \nocite{bww96}
The ``standard'' assumption is that neutron stars are formed  in binaries by 
stars initially 
less massive than 40~\msun\ (van den Heuvel \& Habets 1984). However, 
this limit may be as high as $\sim 60$~ \msun\ 
if helium stars have very strong stellar winds (Woosley \etal 1995).  
On the other hand, there are indications from theory as well as from
observations that this limit may be as low as 20-25\,\msun\ (see 
Portegies Zwart et al.\ 1997b and Timmes et al.\ 1996)
The wide spread in the mass of observed candidate black holes in binaries 
suggests that there may be factors other than the progenitor mass
alone which determines the fate of a star
(Ergma \& van den Heuvel 1997).\nocite{eh97}

The chain of events resulting in the formation of a bound pair of 
neutron stars is known now for more than 20 years, starting from  
studies by van den Heuvel \& Heise (1972), Tutukov \& Yungelson 
(1973), De Loore et al (1975), 
Flannery \& van den Heuvel (1975) and van den Heuvel (1976).
Let us introduce the notation \ms, \gs, \he, \co, \ns\ and \bh, 
for a main-sequence star,
giant, stripped helium core or helium star,  the stripped 
carbon-oxygen core of a helium star, neutron
star and a black hole, respectively 
(see \cite{pzv96} for an extended description). 
A bullet above a stellar type indicates that this
star is to experience a supernova explosion in the next evolutionary
step.
A subscript $e$\ means that the orbit of the system is eccentric.
A bound pair is indicated 
by symbols enclosed by round brackets:
two main-sequence stars in a circular orbit are denoted 
as (\ms, \ms) and an eccentric binary neutron star as (\ns, \ns)\ecc.
A square bracket by the symbol indicates that this star fills its Roche lobe
and stably transfers mass onto its companion.
If the binary experiences a phase of unstable mass 
transfer the symbols are placed within braces.
Using this notation the ``standard'' scenario for the formation of
an (\ns, \ns)\ecc\ is:

\begin{eqnarray*}
\lefteqn{
\hspace*{-0.6cm}{\rm I.}~~(\ms,\ms)\ecc, [\gs,\ms), (\he,\ms), 
       [\he,\ms),(\stackrel{\bullet}{\co},\ms), (\ns,\ms)\ecc,}\\
	\nonumber
\hspace*{-0.6cm}\lefteqn{~~
 \{\ns,\gs\}, (\ns,\he), 
     \{\ns,\he\}, (\ns,\stackrel{\bullet}{\co}),(\ns,\ns)\ecc.}
	\nonumber
\end{eqnarray*}
Actually, all other scenarios for the formation of HMBP
differ from scenario~\I\ in the number of Roche-lobe overflow 
and common envelope 
events experienced by the binary.  
The stages [\he, \ms) and \{\ns, \he\} typically are absent for
helium stars more massive than $\sim 4$ \msun\ which do not expand
much beyond $\sim10\,\rsun$ after core helium exhaustion
(\P\ 1971; Habets 1985; Woosley \etal 1995).  The strong stellar wind of
helium
stars may also prevent the occurrence of both phases of mass transfer from the
helium star.  

A peculiar case are binaries with a primary star initially more
massive than $\sim 40$\, \msun\ and a secondary less massive than 
25 to 40~\msun. 
They may avoid the first \RLOF\ at all due to their strong 
stellar wind which may 
1) prevent them from expanding to a radius
exceeding several hundred solar radii and blow away most of their
envelopes during the Wolf-Rayet phase, and 2) increase the
orbital separation via loss of angular momentum.
However, the secondary star
may overflow its Roche lobe when it evolves onto the giant branch.  The 
orbital period subsequently
reduces as a result of the spiral-in and the second supernova occurs
in a short-period binary.  Note that a kick can also decrease the orbital
separation upon the formation of the first neutron star.

Scenario~\I involves stages when the neutron star is 
immersed into a common envelope (the \{\ns, \gs\} stages and, on some
occasions, also \{\ns, \he\}).
Recently certain doubts were expressed concerning this phase based on 
the work of Zel'dovich \etal (1972, see also Chevalier 1993; 
Brown 1995; Fryer \etal 1996) which have shown that  
neutrino cooling allows hypercritical accretion onto neutron stars, 
which is O($10^8$) higher than the canonical 
photon Eddington limit of $\sim 10^{-8}$~\myr. 
Common envelopes in massive binary systems may 
provide suitable conditions for hypercritical accretion; the net outcome 
may be the collapse of the neutron star into a black hole.
Whether or not the binary survives the common envelope and 
a (\bh, \ns)\ecc\ binary forms or the
spiral-in continues until 
both stars merge into a single object is nor clear.
The radical conclusion (Brown 1995; Wettig \& Brown 1996; Fryer \& 
Kalogera 1997) is that the binaries in which 
neutron star enters a common envelope are not able to produce HMBP.
This reduces the wealth of scenarios for the formation of HMBP 
to the following one: 
\begin{eqnarray*}\lefteqn{
\hspace*{-0.6cm}
{\rm II}.~~(\ms,\ms)\ecc, [\gs,\ms), (\he,\ms),
   \{\he,\gs\}, (\he,\stackrel{\bullet}{\he}), (\he,\ns)\ecc,} \\
	\nonumber 
\hspace*{-0.6cm}\lefteqn{~~~
     (\stackrel {\bullet} {\he},\ns)\ecc,  (\ns,\ns)\ecc.} \\
	\nonumber
\end{eqnarray*}
In scenario \II both binary components must initially have
almost equal mass (within $\sim 4\%$)
and a primary with a relatively shallow convective 
envelope. This allows almost conservative mass exchange in the 
[\gs, \ms) phase. 
Due to the large amount of mass which is accumulated by the secondary star
its evolution raps-up considerably and it 
can overfill its Roche lobe before the 
primary star explodes as a type Ib supernova. 
The first-born \ns\ is subsequently recycled in the
stellar wind of its companion helium star.
The stellar wind prevents the helium common envelope phase.
As the outcome of hypercritical accretion is yet not clear, we 
consider below two options: the formation of a single black hole 
or the formation of a (\bh, \ns) binary for the case where both
components do not merge within the common envelope formalism.

Finally, there exist ``wide'' binaries which avoid mass exchange
altogether.  Both components may evolve independently, if the initial orbital
period is $\ga 10^4$~days or if the initial mass of the components exceeds
30--40~\msun\ and stellar wind drives the starts apart and prevents
\RLOF.  The initial eccentricity is not affected by
tidal effects and it is conserved until the
first supernova event occurs. The
eccentricity increases the survival probability for the binary when
the primary and the secondary subsequently explode in supernovae.  It
is likely that kicks prevent formation of (\ns, \ns)\ecc\ pairs for
(most of the) wide binaries. If one or two are formed anyway, both pulsars
will be young and the binary will be visible as two single pulsars
which happen to be spatially close for about 10 Myr.  Detection of
binarity of pulsars is possible for orbital periods below $\sim
20\,000$ yr (\cite{ll76}). Binarity might be confirmed for very wide
binaries if the proper motion of two stars are the same within 
about a \kms\ (Latham \etal\ 1984).
Detection of a young pulsar in a wide
binary may be an {\em experimentum crucis} for the hypothesis claiming
that single stars and components of wide binaries do not produce radio
pulsars because of the slow spin of the pre-supernova stars (Tutukov
\etal\ 1984; Iben \& Tutukov 1996). It will argue even against
the occurrence of kicks within a few tens of \kms\ 
(Portegies Zwart \etal\ 1997a).\nocite{pzkr97} 

The neutron star in the close binary which was not disrupted by the first
supernova cannot be detected as a radio pulsar due to free-free
absorption of radio emission in the wind of its companion.  
Within 10\,Myr to 20\,Myr, 
before the second supernova occurs, the pulsar passes away.
This
first-born ``dead'' pulsar may be posthumously recycled in the stellar
wind of its companion and/or during the common envelope phase
(Bisnovatyi-Kogan \& Komberg 1975; Smarr \& Blandford 1976; 
Bhattacharya \& van den Heuvel 1991).
The interaction in a binary system is hypothesized to result in a sharp 
decrease of the magnetic field strength down to $\sim 10^9$~G 
(see Bhattacharya 1996 for a list of suggested hypotheses and their 
classification). A weak field results in a prolonged lifetime for the 
recycled pulsar, about two orders of magnitude longer than for a 
``regular'' pulsar born with a high magnetic field and with 
the same pulse period (see \eg\ Fig. 1 in \cite{vdhl96}).  

\subsection{Common envelopes}
The crucial stage in the formation of a HMBP in 
scenario~\I is the common envelope stage
in which the envelope of the 
donor star is expelled from the binary system through the interaction with 
its companion. 
The net result of the common envelope phase is a dramatic reduction in the 
orbital separation. Estimates of this reduction may be obtained 
from a comparison of the binding energy of the common envelope $\Delta 
E_{\rm b}$ with the change 
of the orbital energy of the binary $\Delta E_{\rm orb}$ (see \eg\ Iben \& 
Livio 1993;
Livio 1996 for discussion). The measure of this ratio is the {\em common 
envelope parameter} $\ace = \Delta E_{\rm b}/\Delta E_{\rm orb}$. 
Tutukov \& Yungelson (1979) 
suggested the following equation for the variation of the separation of 
binary components in the common envelope (the donor mass is $M$ and
its companion has mass $m$):
\begin{equation}
\frac {(M_{\rm i}+m)(M_{\rm i}-M_{\rm f})}{2a_{\rm i}} = \ace 
M_{\rm f} m \left( \frac {1}{2a_{\rm f}} - \frac{1}{2a_{\rm i}}\right),
\label{eqty}\end{equation}
where subscripts $i$ and $f$ refer to the initial and final states, 
respectively.
This equation is based on the assumption that the 
spiral-in starts when the envelope 
surrounding the two cores has dimension of $\sim 2 a_{\rm i}$ 
and the envelope is ejected from the gravitational potential of 
both components. 
Recent 3-D \SPH\
calculations for the initial stage of formation of common envelopes 
(\cite{rl96}) confirm such an assumption, at least qualitatively, by 
showing that very little mass is expelled during this stage and that 
the envelope expands considerably.

On the other hand, Webbink (1984) suggested a more 
straightforward equation for $a_{\rm f}/a_{\rm i}$ variation:
\begin{equation}
\frac {M_{\rm i}(M_{\rm i}-M_{\rm f})}{a_{\rm i} \lambda r_{\rm L}} =
\eta_{\rm ce} m \left( \frac {M_{\rm f}}{2a_{\rm f}} - 
\frac{M_{\rm i}}{2a_{\rm i}}\right), 
\label{eqw}\end{equation}
where $\lambda$ is a structural constant 
which depends on the density 
profile in the star [$\lambda = 3/(5-n)$ for polytropical models with
index $n$; the usual assumption is $\lambda = 0.5$], $\eta_{\rm ce}$ is  
the common envelope parameter and $r_{\rm L}$~ is the fractional 
radius of the Roche-lobe of the donor. Actually we handle the product
\elce\ as a single parameter.

For completely non-conservative common envelopes, 
both equations result in 
equal $a_{\rm f}/a_{\rm i}$ for $\ace \approx 4 \eta_{\rm ce} \lambda$.
Confrontation of population-synthesis calculations with observations
does  not 
provide very rigorous constraints on the common envelope parameter.
Nevertheless, population synthesis of
galactic symbiotic stars (Yungelson \etal 1995), 
Be/X-ray binaries (Portegies Zwart 1995), 
supersoft X-ray sources (Yungelson \etal 1996), 
close binary white dwarfs (Iben \etal 1997) and
low-mass millisecond pulsars (Tauris 1996; Tauris \& Bailes 1996)
confirms that using a high value of the common-envelope 
parameter gives a better agreement with the observations.
Van den Heuvel (1994) finds a reasonable
evolutionary scenario with $\eta_{\rm ce}\lambda = 2$ for {\mbox PSR~ 
J2145-0750}, a low mass binary pulsar, 
while for $\eta_{\rm ce} \lambda = 0.5$~ the spin-up of the neutron star 
in this particular case is problematic. 
Increasing $\eta_{\rm ce} \lambda$ to values larger than unity
decreases the discrepancy between the predicted and 
observed ratios of the birthrates for neutron star and black hole 
low-mass X-ray binaries (see Fig.~4 in \pz\ \etal\ 1997b).
On the basis of 3-D \SPH\ calculations \cite*{rl96} claim
that the deposition of energy into the common envelope must be 
highly efficient.
Taken at face value, for Eq.~\ref{eqw} values of $\eta_{\rm ce} >1$~mean 
in that additional energy 
for the expulsion of the common envelope must come from sources other than 
the orbital energy. However, Eqs.~\ref{eqty} 
and \ref{eqw} are by no means anything more 
than indicative order of magnitude estimates.
We argue below that reproduction of 
the distribution function for: the orbital period, the 
eccentricity and the relative birthrate of 
observed HMBP are reconciable 
with high values for the common envelope parameter.     

\subsection{The model} 
For the present study we used the population
synthesis code {\sf SeBa} described by Portegies Zwart \& Verbunt
(1996)\nocite{pzv96}.
The basic difference with the Tutukov \& Yungelson (1993a,b) code is 
that they use Eq.~\ref{eqty} instead of Eq.~\ref{eqw} which is used
by Portegies Zwart \& Verbunt for
the treatment of binaries which evolve non-conservatively in the 
semi-detached and common-envelope phases. 
A detailed comparison of evolutionary sequences for models with similar
initial parameters (\cite{pzv96}) shows that the sets of assumptions
in the two codes
result in reasonably similar configurations prior to
the second supernova explosion in the system (although the stages in
between can be quite different). 
Moreover, the
enhancement of the amount of angular momentum loss during
non-conservative mass transfer in the present model [$\eta_J=6$
instead of $\eta_J=3$, see Eq. (38) in Portegies Zwart \& Verbunt
(1996)],
 makes the results
more comparable\footnote{For binaries with a mass ratio below $\sim
0.4$ assumption of $\eta_J = 6$ is similar to mass loss through the
second Lagrangian point, for larger mass ratios less angular momentum
is lost.}.
Comparison of synthesis studies with the
observed population of Be/X-ray binaries supports this choice for loss
of angular momentum (Portegies Zwart 1995).
For the present study the mass-loss
law in the stellar winds from the helium stars was accepted 
after Langer (1989): $\md = 5 \times 10^{-8} M^{2.5} \myr$, which is
about twice the value used by \cite*{pzv96}.

After the initial conditions for each binary are selected it is
evolved until the binary merges, is disrupted or a pair of neutron
stars is formed.  For the latter, further decay of the orbit is
computed from linearized general relativity (Peters
1964).\nocite{pet64} At each moment in time, the distributions of the
orbital elements are computed assuming a continuous formation of new
systems and taking into account the decay of the orbital elements of the
present (\ns, \ns)\ecc\ binary population.

The zero-age parameters of the simulated binary population are
selected from independent distribution functions.
The mass of the primary $M$ is chosen from a mass function of the form:
\begin{equation}
	dN \propto M^{-2.5}dM.
\label{eqimf}\end{equation}
The mass of the secondary star $m$ is selected from a distribution
function for the mass ratio $q_o \equiv m/M:$ 
\begin{equation}
\Psi(q_o) \propto (1 + q_o)^\phi.
\label{eq:qdist}\end{equation}
We explored the dependence of the model upon $\phi$ in the range from -4 
to +5 and for a ``flat'' distribution $\Psi(q_o) = 1$.
For the semi-major axis distribution we explored two options which 
are consistent with observational data: flat in $\log a$ between 
0.1~\rsun\ and $10^6~$\rsun\ (Abt 1983) and a Gaussian 
with a mean of $10^{3.9}$\rsun\ and a dispersion of $10^{2.9}$\rsun\ 
(Duquennoy \& Mayor 1991).
The eccentricity for a binary was chosen between 0 and 1 from the
thermal distribution: $\Xi(e) = 2 e$.

Observations of the space velocities of single radio pulsars suggest that
neutron stars may receive a kick at birth (Gunn \& Ostriker 
1970).
A clear summing up of arguments for asymmetric kicks  is
given by van van den Heuvel \& Paradijs  (1997)\footnote{See, 
however, Iben \& Tutukov (1996) for arguments {\it contra} kicks.}. 
Following Hansen \& Phinney (1996) and Hartman (1997)  we used  the
distribution  
for isotropic kick velocities (\P\ 1990)
\begin{equation}
P(u)du = {4\over \pi} \cdot {du\over(1+u^2)^2},
\label{eqkick}\end{equation}
with $u=v/\sigma$ and $\sigma = 600~\kms$.
Implementation of this distribution of kicks by Hansen \& Phinney and 
Hartman 
provides a reasonable model for the population of single pulsars close
to Sun.  As an alternatives we
also made computations without kicks and with a Maxwellian
kick-velocity distribution.

All stars are supposed to be born in binaries.  The total birthrate is
normalized to the present astration rate of~ \mbox{4\,\myr,} together with
$M_{\rm min} = 0.1$\,\msun\ and Eq.~\ref{eqimf}, which is consistent with
observational estimates for the Galactic disc (see van den Hoek 1997;
van den Hoek \& de Jong 1997).  The inaccuracy in the number of
binaries formed with $M_1 \geq 8 \msun$ due to uncertainties in the
astration rate and the shape of mass-function below \mbox{$\sim 0.3 \msun$}
may be O(2).  The assumed age of the Galactic disc is 
$T = 10$\, Gyr\,
(Meynet \etal 1993; Carraro \& Chiosi 1994). The star formation
rate is assumed to be constant throughout the evolution of the Galaxy.
For all models a total of $5 \times 10^5$ binaries 
with a primary mass between 8\,\msun\ and 100\,\msun\ are initialized.

\section{Results}
Our main purpose is to model the present Galactic population of (ns,
ns)\ecc\ binaries with the aim to reproduce their birthrate,
period-eccentricity distribution and to obtain an estimate for their
merger rate and extrapolate it to the local Universe.

\subsection{Birth rates of neutron stars and related binaries}

In each of the eight models, for which we present the results below, different
assumptions are made concerning the kick velocity, the common envelope
parameter, the mass ratio and semi-major axis distributions.
Basic peculiarities of the models are summarized in Table~\ref{tabbr}.
As a ``standard'' set of assumptions we consider the one with a flat initial
distributions in $q_o$~ and $\log a$, common envelope parameter
$\elce = 2$, and the kick distribution given by
Eq.~\ref{eqkick}.

In model A no kicks are imparted to neutron stars or black holes.
 
Model B is identical to model A, but now neutron stars do receive a
velocity kick upon birth chosen randomly from the distribution 
given by Eq.~\ref{eqkick}. This model is ``standard''. Black holes also 
receive a kick, but 
scaled by mass:~ $v_{\rm bh} = v_{\rm ns} m_{\rm ns}/m_{\rm bh}.$ No mass is 
ejected in the collapses to black holes, contrary to neutron stars. 
Small, but not vanishingly, kicks for 
black holes may be consistent with the velocity dispersion of 
$\sim 40\,\kms$~of X-ray transients with black-hole candidates (White \& van 
Paradijs 1996)  
which is higher than velocity dispersion of O-stars ($\sim 10 - 20\, \kms,$~ 
Wielen 1992).

Model C 
is the only model in which the kick velocity is selected from a Maxwellian
distribution with a 3-dimensional velocity dispersion of $\sigma = 450\, \kms$.

Model D uses a smaller value for the common envelope parameter of
$\elce = 0.5$.

The initial semi-major axes of orbits in model E are selected from a 
Gaussian (see sect.~2.3).

In the models F and G the mass-ratio distribution is not flat:
$\phi = -4$~in model E and +5 in model G (see Eq.~\ref{eq:qdist}).
 
Finally, in model H which used the standard assumptions of model B, 
the accretion rate onto a \ns\ in a common envelope is 
$10^8 \times \md_{\rm Edd}$.  
All neutron stars which are able to increase their mass to $\ge 2$~ 
\msun\ are to become black holes. No kicks are applied in these cases.

\begin{table*}
\caption[]{
Results of the computations for a selection of the computed models.
Columns (2) to (5) give  
the birthrates of single young radio pulsars,
binaries which contain a
young neutron star or a black hole and a non-compact star, (\ns, 
$\star$) and (\bh, $\star$), respectively, and
binaries which contain a black hole and a young neutron star (\bh, \ns).
Column (6) gives the birthrate of single recycled
pulsars, column (7)  -- the occurrence rate of mergers between neutron 
stars and non-compact companions \{\ns, $\star$\} 
and column (8) gives the birthrate of HMBPs: (\ns, \ns). 
Columns (9) and (10) give the merger rate of neutron star binaries 
\{\ns, \ns\} and binaries that contain a neutron star and a black hole
\{\bh, \ns\}, after $10^{10}$ years of the evolution of the Galaxy 
with a constant star formation rate.  Birthrates are normalized 
to galactic astration rate of 4\myr\ and 100\% binarity.
The numbers in parentheses for model H give the rates for the 
case when hypercritical accretion does not always 
result in the merger of the components.
The statistical (Poissonian)
uncertainty in the rates is smaller than 1\%. }
\begin{flushleft}
\begin{tabular}{l|rrll|rrr|rl|l}
\hline\noalign{\smallskip}
 & \multicolumn{7}{c|}{birthrate$\times 10^5\, \pyr$} 
 & \multicolumn{2}{c|}{merger rate$\times 10^5\, \pyr$}
 & comments \\ \hline
& \multicolumn{4}{c|}{young pulsars} 
 & \multicolumn{3}{c|}{recycled pulsars}
 & & \\ 
& ns &(ns,$\star$) &(bh,$\star$) & (bh,ns)
 & ns & \{ns,$\star$\} & (ns,ns) 
 & \{ns,ns\} & \{bh,ns\} & \\ 
(1) & (2) & (3) & (4) & (5) & (6) & (7) & (8) & (9) & (10) \\ \hline
A & 940 & 500& 72 & 18.7
        &  183.5 & 1.5 & 60.0 
	&   39.0 & 0.0 & no kick \\ 
B & 1470  & 60 & 7.6  & 0.6 
        &   15.4 &  7.0 & 3.4 
	&    2.3 &  0.1 &  standard model; kick from Eq.~5 \\
C & 1490  & 40 & 5.4 & 0.4 
	&    6.9 & 9.7 & 1.7 
	&    1.2 &  0.1 &  kick from Maxwellian distribution \\
D & 1480  & 53 & 7.5 & 0.8
      	&   11.9 & 12.6 & 0.8 
	&    0.3 & 0.2 &  more dramatic spiral-in: $\elce = 0.5$\\
E & 1440  & 69 & 8.7 & 0.8
	&   16.4 & 8.8 &  3.8 
	&    2.6 & 0.1 &  Gaussian major axes distribution \\
F & 1490 & 52 & 7.4 & 0.7
	&    5.9 &  6.4 & 1.7 
	&    1.3 &  0.06 &  mass-ratio distribution peaked to 0 \\
G & 1510 & 66 & 6.9 & 0.2
	&   29.3 & 5.2  & 5.3  
	&    3.4 & 0.04 &  mass-ratio distribution peaked to 1 \\
H & 1420 & 57 & 7.2 & 0.6 (5.3)
	&   10.6 & 5.7 & 0.7 
	&   0.2 &  0.1(3.5) &  Hypercritical accretion onto \ns \\ 
\end{tabular}
\end{flushleft}
\label{tabbr}
\end{table*}

Table~\ref{tabbr} gives an overview of a selection of the computed
models.  The sum of the entries in the columns~(2) to (5), and (8) gives
for each model the annual rate of explosive events resulting in the
birth of a young neutron star or a black hole.  The estimated rate of
formation of {\em single} neutron stars is given by the sum of the entries
in columns~(2), (6) and (7). In all models these rates are close to
$1.5 \times 10^{-2}$~\pyr\ and are almost independent of the
assumed semi-major axes and mass-ratio distributions of the binaries
and the rate of binarity.  This can be understood as a consequence of
the relatively small number of high-mass secondaries.  A smaller
binary fraction together with the same astration rate results in an
increase of the number of single stars and therefore a higher
birthrate for single pulsars.  If all stars are born single the pulsar
formation rate becomes $1.9\times 10^{-2}$~\pyr.  The model birthrate
of pulsars may be compared to the inferred Galactic occurrence rate of
type II and Ib/c supernovae: $(12\pm6) \times 10^{-3}\,\pyr$ for SN~II and
$(2 \pm 1) \times 10^{-3}$\,\pyr\ for SN~Ib/c 
(Cappellaro \etal~1997)\footnote{For the
morphological type $Sb$, blue luminosity of~ $2 \times 10^{10} \lsun$
and $H_o = 75$ \mh.}.  Based on the modeling of population of pulsars
close to Sun, Hartman \etal (1997a) derive a pulsar formation rate of
$(9 \pm 3) \times 10^{-3}$~\pyr, where a beaming factor of 0.3 was
assumed.

The relative birthrate of single and binary neutron stars is most
strongly affected by the assumptions concerning kicks 
(compare model A with the
models B to H, see Table~\ref{tabbr}).  The formation rate of single
non-recycled neutron stars, column~(2) in Table~\ref{tabbr}, increases
considerably when kicks are applied. This is not surprising because
the majority of the wide non-interacting binaries and a considerable
fraction of the binaries with shorter orbital period that experience
a rather stable phase of mass transfer, which causes the orbital
separation to increase, are dissociated upon the first supernova. On
the other hand, the proportion of binaries which survive the first
supernova explosion depends only weakly on the initial distributions
for the mass ratio, semi-major axis, and the shape of the
kick velocity distribution. The reason is that under any reasonable
$\Psi (q_o)$ and $\phi (a)$ most of the close binary systems evolve to
shorter periods and then it is mainly the average $v_{\rm kick}$ which
determines 
the survival probability in the supernova explosions (see \eg\ Fig.~8
in Lipunov \etal 1996).
   
The formation rate of bound neutron stars, (ns, $\star$), column~(3),
drops by an order of magnitude when kicks are present.  This entry in
Table~\ref{tabbr} gives a rough estimate of the formation rate of
X-ray binaries. Since the majority of them are high-mass binaries,
their total number in the Galaxy can be estimated.  Actually, these
systems generally start as a Be/X-ray binary to evolve into an ordinary
high-mass X-ray binary at a later instant.  The X-ray lifetime of a
Be/X-ray binary is between 1 and 10 Myr (the remaining main-sequence
lifetime of the B-star companion of the neutron star).  The estimated
galactic population is consequently somewhere between $10^3$ and
$10^4$.  The lifetime of high-mass X-ray binaries is of the
order of $10^3$\,yr to $10^5$\,yr, depending on the evolutionary
status of the donor (e.g.\ Savonije 1979; Massevich \etal 1979;
Hellings \& DeLoore 1986).  This results
in a total of several tens of high-mass X-ray binaries in the Galaxy.
These estimates of numbers of Be- and high-mass X-ray sources are
consistent with observational estimates of $>2000$ and $55 \pm 27$,
respectively (Meurs \& van den Heuvel 1989, and 50 to 80 by Iben
\etal\ 1995). Models without a kick
seem to overestimate the numbers of X-ray binaries by an order of
magnitude.

Binaries which contain a black hole and a stellar companion are not as
common as binaries which contain a neutron star.
If their lifetime is comparable to the lifetime of high-mass X-ray
binaries which contain a neutron star,  
their number has to be $\sim 1/10$ of the former,
in a agreement with the observations. 

Derivation of the expected birthrate and the number of low-mass X-ray
binaries from our calculations would require a more rigorous
consideration of very severe constraints on the orbital periods and
masses of donors in prospective candidates (Kalogera \& Webbink 1996)
and is beyond the scope of the present paper.

Entries in columns~(6) to (8) of Table~\ref{tabbr} refer to neutron
stars which were born first in the binaries and were subsequently
recycled or merged with their stellar companions.  Evidently, the
interaction with companions influences the radio properties of neutron
stars.  We still call them ``recycled'' pulsars, though decay of the
magnetic field and/or the action of the propeller mechanism can
prevent a certain fraction of them to show up in radio.
The birthrate of single recycled pulsars is given in column~(6) of
Table~\ref{tabbr}. Observational properties of these objects are
discussed in detail by Hartman et ~al.\ (1997b). 
Typically, single recycled pulsars contribute for less
than 1\% to the total population of single radio pulsars\footnote{The
birthrates found by Hartman \etal are smaller by about a factor two
because they assumed a binarity fraction of 50\%.}.  This is
consistent with the presence of eight isolated objects among observed
millisecond pulsars in the Galactic disc.  
In the absence of kicks
(model A)~ $\sim 1/7$ of all pulsars are old and possibly
recycled; this contradicts the observations 
(see however Deshpande \etal 1995 for arguments which suggest a high
fraction of recycled pulsars).

The presence of kicks significantly increases the proportion of
systems in which the neutron star merges with its stellar companion.
The main reason for this is that a kick can decrease the orbital
period or even shoot the neutron star directly into its companion
(see also Brandt \& Podsiadlowski 1995). 
In the first case, the smaller orbital separation of the binary causes a
merger to occur more easily after a phase of unstable mass transfer.
If the merger does not lead to the formation of a black hole the
neutron star may, after ejection of the common envelope, emerge as a
single recycled pulsar.  These systems may increase the fraction of
recycled pulsars by a factor of two.

\subsection{Neutron star--neutron star binaries}

Comparison of column~(3) with column~(8) from Table~\ref{tabbr} shows 
that less than $\sim 5\%$ of the Be- or high-mass X-ray binaries 
survive the second supernova and produce a recycled binary pulsar, 
about 25\% is dissociated in the second supernova and about 10\%
merge into a single object. The remaining $\sim 60$\% never
experience the second supernova. A minor proportion of the latter systems 
may become LMXB and, later,  millisecond pulsars.

We confirm the result from previous computations that
the birthrate of (\ns, \ns) binaries drops by
more than an order of magnitude when kicks are implied.
This reflects in the birthrates in Table~\ref{tabbr}.
More binaries are dissociated when the kick velocity is taken from a
Maxwellian distribution instead of the distribution of Eq.~\ref{eqkick},
because of the larger average magnitude of the kick. 

Directly after the second supernova, a binary containing two
neutron stars of which one is visible as a young radio pulsar and the
other may be recycled. The later born pulsar dies after about 10~Myr. 
A recycled pulsar may live up to a few billion years. For most of
this time span the binary may be detectable as a
recycled pulsar with a ``dead'' companion.
A few known systems are listed in Table~\ref{tabobserved}.
For PSR~J1518+49 and B2303+46 only the total mass
is known and they may have a massive white dwarf instead of a neutron
star as a companion.
White dwarf companions will be ruled out if optical 
observations will not discover objects of a visual magnitude of 
$\sim 26$ for PSR~J1518+49 and between $23$ to $25$ for the other pulsar
(Fryer \& Kalogera 1997).

\begin{table*}
\caption[]{Observed population of massive neutron star binaries. Their names
are followed by the pulse period, the period
derivative and the orbital elements, age and magnetic field strength.
References: 1 - Nice et al.\ (1996), 
            2 - Wolszczan (1990), 
            3 - Lyne \& McKenna (1989), \cite{bh91b}, 
            4 - Hulse \& Taylor (1975), 
            5 - Deich \& Kulkarni (1996), 
            6 - Stokes et al.\ (1985). 
Data which is not available in the cited papers were taken from Taylor
\etal (1993).  
}
\begin{flushleft}
\begin{tabular}{ll|rrrr|rr|l}
\hline\noalign{\smallskip}
PSR      & ref. & $P$    & $\log \dot{P}$& 
$P_{orb}$ & e     & age & $\log B$ \\
         && [ms]   & [${\rm s\,s^{-1}}$]      &  days  &      &[Myr]& [G]  &
Comments \\ 
\hline
J1518+49 &1&  40.94 &-19.4 &  8.634 & 0.249 &16000& 9.1     &  \\
B1534+12 &2&  37.90 &-17.6 &  0.420 & 0.274 &250 & 10.0   & \\
B1820-11 &3& 279.83 &-14.86&357.762 & 0.795 &3.3& 11.8 &  
Non-recycled? \\
B1913+16 &4&  59.03 &-17.1 &  0.323 & 0.617 &110 & 10.4 & \\
B2127+11C&5&  30.53 &-17.3 &  0.335 & 0.681 &100 & 10.1 
& In globular cluster M15    \\
B2303+46 &6&1066.37 &-15.24& 12.340 & 0.658 &30& 11.9 & 
Non-recycled? \\ \hline
\end{tabular}
\end{flushleft}
\label{tabobserved}
\end{table*}

The long pulse period and strong magnetic field of PSR~B2303+46 do not
exclude that it is the second born pulsar which we observe.  In this
case, its companion would be at most $\sim 2\times 10^7$ yr older (the
evolutionary timescale of a $M \sim 10\, \msun$~ star).  
The companion is either a dead pulsar which did not resurrect as a
recycled pulsar or it is beaming away from our planet.  This discussion
also applies to PSR~B1820-11 for which even the total mass is not
known.  It might be the young pulsar with a main-sequence star, a
white dwarf or an unseen pulsar as companion 
(see \cite{pv91} and \pz\ \& Yungelson (1997), for discussion).

The birthrate of radio pulsars in neutron star binaries must equal the
birthrate of recycled pulsars in similar systems (Bailes 1996).  The
absence of ``young'' pulsars in pairs with old ones among $\sim 650$
objects restricts the {\em relative} birthrate of (\ns, \ns) systems
to 1/650.  Bailes' criterion does not restrict the {\em absolute}
birthrate of (\ns, \ns) binaries; it lacks absolute calibration.
Except for model~A, none of the models contradicts this statement.  If
it is actually the young pulsar we observe in PSR~B2303+46 and
B1820-11 the limit on the relative birthrate increases to 2/650.

\subsection{Orbital period--eccentricity plane for neutron-star binaries} 

\begin{figure}
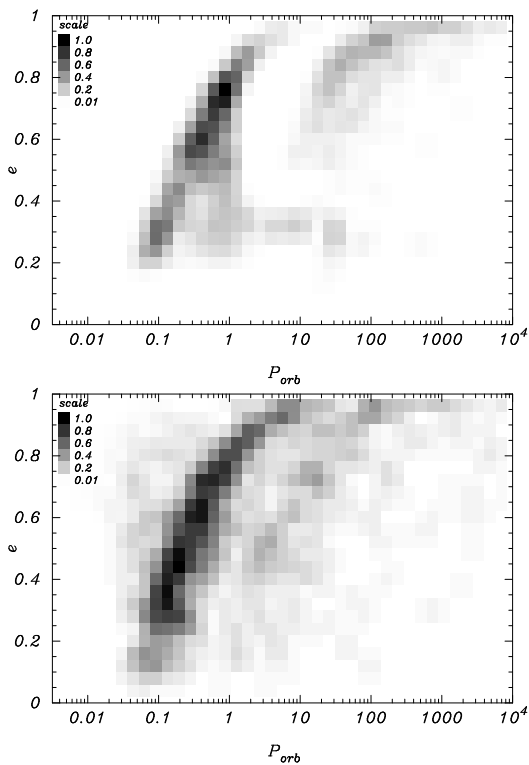

\centerline{
\psfig{file=a4q-1f0k_T0.ps,bbllx=575pt,bblly=40pt,bburx=95pt,bbury=705pt,height=5cm,angle=-90}}
\centerline{
\psfig{file=a4q-1fpp_T0.ps,bbllx=575pt,bblly=40pt,bburx=95pt,bbury=705pt,height=5cm,angle=-90}}
\caption{
The probability distribution for the orbital parameters 
of the (\ns, \ns) binaries with a
period smaller than $10^4$ days
at the moment of birth.
The upper panel gives the probability distribution for the model
with $\elce = 2$, flat initial distributions over $q_o$ 
and $\log a$ and without a 
kick; lower panel is for the same model with a distribution of kicks
according to Eq.~5. 
Darker shades indicate higher
probability to find a neutron star binary at that location in the
period-eccentricity plane. 
The gray scaling is given in the upper left corner of each figure, the
darkest shade in the upper panel
corresponds to a birthrate of $1.2 \times 10^{-5}$\, \pyr, for the
lower frame this value is $3.5 \times 10^{-7}$\,\pyr.
The integrated birthrates are given in Table~1, column~(8)
}
\label{figzero_age}
\end{figure}

The only observational parameters of HMBPs
to which one may compare the results
of evolutionary computations and population synthesis are the orbital
period $P_{orb}$ and eccentricity $e$.  Other characteristics of the
binary like the center-of-mass velocity are ill known from 
observations.  For a comparison of the characteristics of the
individual pulsar, like the strength of the magnetic field, the
lifetime and the pulse period, no theory is available.  In the
\pe~diagram it is only the emission of gravitational waves that
effects the further evolution of the orbital parameters. 

Figure~\ref{figzero_age} gives the distribution function for the
orbital parameters of (\ns, \ns) binaries at birth for a model without
a kick (A) and for model~B with a kick velocity distribution according
to Eq.~\ref{eqkick}.  The population of (\ns, \ns) binaries with
periods beyond $10^4$ days is not shown. 
Since in models without kick most wide
binaries survive the supernovae explosions bound, $\sim 70 \%$\ of all
(ns,ns) have orbital periods greater than $10^4$\, day. Nondetection
of young pulsars in wide binaries then argues for kicks of at least of
several tens of \kms, see Portegies Zwart \etal\ 1997a. 
In the models 
with a kick this population contributes only little ($\la 10\%$, see also
\pz\ \& Verbunt 1996).  In both models the adopted common envelope
parameter in Eq.~\ref{eqw} is $\elce = 2$ (which is roughly equivalent
to \ace = 0.5 in Eq.~\ref{eqty}).  The reason for this choice is the
inability for smaller $\elce$~ to reproduce the position of the observed
high-mass recycled pulsars in the \pe~plane.  Particularly, the
probability for the formation of binary pulsars with $P_{orb} \apgt
2$~ days and $e \aplt 0.3$ and $P_{orb} \sim 10$~ days and $e \sim 0.6
- 0.7$ appears to be very small.  The reason is that low efficiency
of the deposition of orbital energy into the common envelope results
in the formation of very close binaries.  For comparison we give the
\pe\ diagram for $\elce = 0.5$ in Fig.~\ref{figeta1}.

If no kicks are imparted the \pe\ diagram has clear-cut borders (see
Fig.~\ref{figzero_age}).  The shortest orbital period is determined
by the requirement to accommodate a helium star with a neutron star
companion in the pre-explosion orbit. The supernova tends to widen the
binary orbit and the minimum eccentricity is obtained from the
fact that at the instant of explosion all pre-neutron stars still have
helium envelopes of $\Delta M \apgt 0.5 \msun$.
A minor
population with low $e$ and $20 \aplt P_{\rm orb} \aplt 30$ day
descends from massive systems which had eccentric orbits before the
second explosion.  The major, short-period, population in Fig.~
\ref{figzero_age} descends from systems which evolved via scenario~\I
with all mass exchanges being unstable, while wider systems
had at least one phase of stable mass exchange before the formation of
the first neutron star. Presence of these two populations was also
clearly seen in the figures in \TY\ (1993a) and \pz\ \& Spreeuw (1996)
papers.
The lower panel in Fig.~\ref{figzero_age} demonstrates
that, apart from reducing the birthrate, kicks have the effect of
``smearing'' the period--eccentricity distribution of the neutron star
pairs. The eccentricity distribution is more strongly affected by the
presence of a kick than the period distribution.  Striking in the lower
panel in Fig.~\ref{figzero_age} is the existence of a population of
binary neutron stars with orbital periods below $\sim 0.1$ day, in
contrast to the computations without a kick.  These short periods are
the result of a kick in the ``right'' direction to reduce the orbital
period upon the second supernova. Due to the emission of gravitational
waves this population is extremely short-living.  If kicks are absent,
only the Hulse-Taylor pulsar is born in the region in the \pe diagram
with high formation probability.  The extremely small probability for
the formation of zero-age pulsars with parameters similar to that of
J1518+49 and B2303+46 is appealing.

\begin{figure}
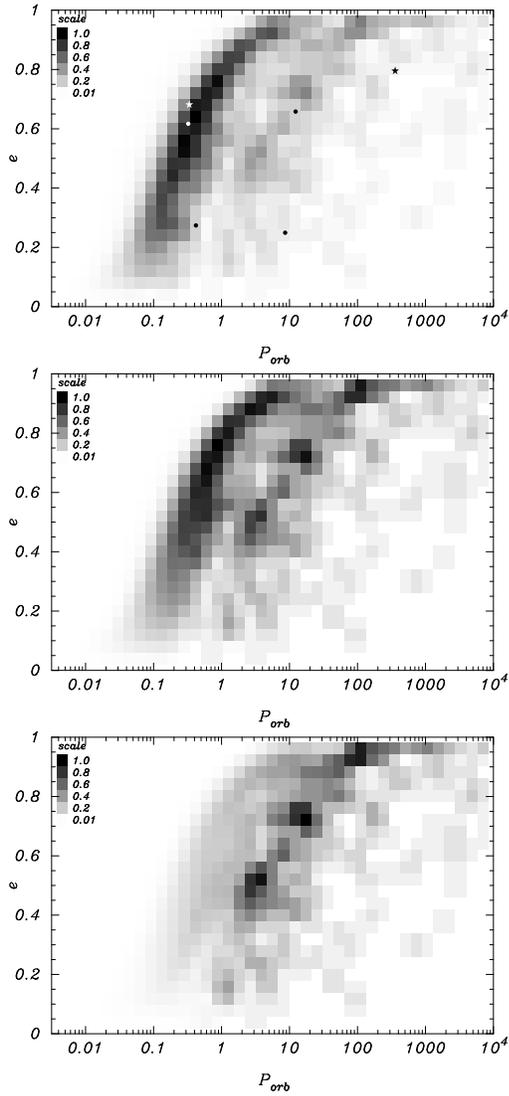

\centerline{
\psfig{file=a4q-1fpp_T2.ps,bbllx=575pt,bblly=40pt,bburx=95pt,bbury=705pt,height=4.8cm,angle=-90}}
\centerline{
\psfig{file=a4q-1fpp_T3.ps,bbllx=575pt,bblly=40pt,bburx=95pt,bbury=705pt,height=4.8cm,angle=-90}}
\centerline{
\psfig{file=a4q-1fpp_T4.ps,bbllx=575pt,bblly=40pt,bburx=95pt,bbury=705pt,height=4.8cm,angle=-90}}
\caption{ The probability distribution in model B for the present day
orbital parameters of the Galactic disc neutron star binaries younger
than 100\,Myr (upper panel), 1\,Gyr (middle panel) and 10\,Gyr (lower
panel).  The gray scaling in this graph represents numbers in the
Galaxy.  The darkest shades in the upper, middle, and lower panels
correspond to a total of Galactic 23, 110, and 1100 (\ns, \ns)
binaries with given combination of $P_{orb}$~ and $e$.  The
``$\bullet$'' symbols in the upper panel represent the observed (\ns,
\ns) binaries, the ``$\star$'' symbols mark the position of the a
system which might originate from a dynamical encounter (the left
$\star$) and a system of which one companion is possibly not a neutron
star (see Table~2).
}
\label{figPPkick}
\end{figure}

Figure~\ref{figPPkick} shows the effect of time evolution on the
population of neutron star binaries from the standard model~B with a
kick, giving the probability distribution of them at the age of
100\,Myr, 1\,Gyr and 10\,Gyr. The limit of 100\,Myr is suggested by
the characteristic age of PSR~B1913+16.  Figure~\ref{figPPkick}~may
be compared with the lower panel of Fig.~\ref{figzero_age} which
gives the probability distribution of these binaries at birth (which
is more appropriate for relatively young pulsars like PSR B2303+46).
PSR~B1913+16 fits extremely well into the model population for its
age.  The rest of the object fall into regions with probability by a
factor of a few lower, but still relatively high. The observed system
may be found in the short period area of the diagram if the birthrate
of its progenitors is high or if the pulsar is old.  The first is
probably the case for PSR~B1913+16 and the other option is for older
PSR~B1534+12. Pulsar PSR~J1518+49 is found in a low-birthrate area of
\pe\ diagram which suggests a long age.  Relatively young pulsars,
like PSR~B2303+46, may be found in any area of the \pe\ diagram except
for the very short periods where the merger time is smaller than the
characteristic age.

The probability distribution shifts to larger orbital period as the
population becomes older, because short-period objects merge due to
the influence of gravitational wave radiation while the long-period
binaries survive for a longer time.  As the age of the population
increases pulsars older than PRS~B1913+16 are found in the area of the
\pe\ diagram with a higher probability; the former pulsar drifts to
the lower probability area.  The middle panel of Fig.~\ref{figPPkick}
fits best with the observed four systems suggesting that the average
age of the observed population is close to 1\,Gyr.  It is, however,
hard to quantify this statement. Particularly, there may be a
differential selection effects related to the different lifetimes of
recycled pulsars, which depends on their specific accretion history.

The bottom panel of Fig.~\ref{figPPkick}~ represents the {\em total}
population of binary neutron stars currently present in the Galaxy.
The (\ns, \ns) binaries we observe today are a  subset of this
population.  Since typical merger time for most (\ns, \ns) binaries is
several hundred Myr (see Fig.~\ref{figage} lower panel) and radio lifetime of
recycled  pulsars is $\aplt 1$\, Myr (van den Heuvel \& Lorimer 1996),
the majority of the neutron star
binaries present in the Galaxy are dead! 

\begin{figure}
\centerline{
\psfig{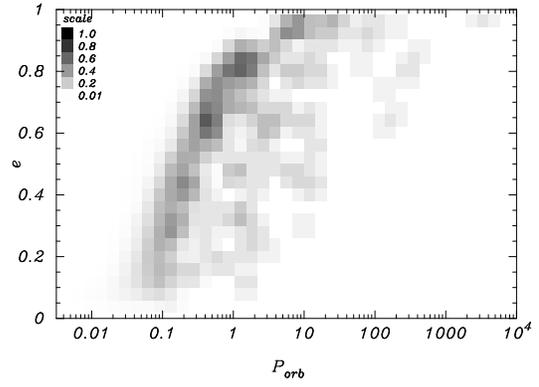}}
\caption{
The probability distribution for the number of Galactic (\ns,~\ns) 
binaries younger than 100\,Myr.
The kicks are selected from a Maxwellian velocity distribution (model C 
in Table~1).
The gray scaling is given in the upper left corner and is chosen to be
the same as for the upper panel in Fig.~2.
}
\label{figMaxwell}
\end{figure}

Figure~\ref{figMaxwell} gives the probability distribution for (\ns,
\ns) pairs younger than 100\,Myr in the \pe\ plane for the model where
the kicks are selected from a Maxwellian velocity distribution.  The
gray shading in this Figure is normalized to that in the upper panel
of Fig.~\ref{figPPkick}. This model has a smaller (\ns, \ns) birthrate
than model B, this reflects in the absence of the darkest shades in
Fig.~\ref{figMaxwell}.  The presence of a non-negligible, yet not
observed, population of (\ns, \ns) binaries with orbital periods
larger than $\sim 10$~days in model~B (see Fig.~\ref{figPPkick})
reflects the higher probability of low-velocity kicks in the
distribution given by Eq.~\ref{eqkick} relative to the Maxwellian.
Though the latter model suggests almost equal probability for the
detection for PSR~B1534+12, B1913+16 and B2303+46, it gives an
extremely low birthrate for binaries with orbital parameters similar
to PSR~J1518+49.  Even high age does not increase the model galactic
population of pulsars with similar orbital parameters.  This
discrepancy does not go away when the velocity dispersion of the
Maxwellian distribution is decreased by a factor of 2.  (Hansen \&
Phinney 1997, suggested that a Maxwellian kick-velocity distribution
with $\sigma = 190\, \kms$\ fits satisfactory with the observed single 
pulsar velocities.)

Model E, with initial Gaussian distribution over $\log a$ 
differs only slightly from the model with an initial flat distribution in
$\log a$. It 
gives a similar birthrate of (\ns, \ns) binaries and reproduces the
\pe\ relation equally satisfactory. 
This means that most of (\ns, \ns) stars descend from 
initial systems having orbital separations in the interval where the
probabilities for formation for both distributions are comparable.   

Our computations included several additional runs with different
parameters which produced models which we consider less satisfactory
than the standard model.  Model F with initial $q$-distribution
strongly raising to 0~ ($\phi = -4$, in Eq.~\ref{eq:qdist}) gives
birthrates of single neutron stars and (\ns, \ns) binaries similar to
these in the standard model B, but does not reproduce the \pe\
distribution satisfactory.  Particularly, it does not reproduce the
population with $P_{orb} \sim 10$~ day and $e \aplt 0.7$ and pulsars
with $P_{orb} \sim 0.4$~ days and $e \approx 0.2 - 0.3.$~ This is an
understandable result of the high proportion of mergers in the first
(hydrogen) common envelope event and of strong reduction of the
orbital separation in the systems which avoid mergers.  Model G, with
$q_o$ strongly raising to 1~ ($\phi = +5$), does not basically differ
from the other models in the total production rate of pulsars and it
reproduces the \pe\ relation rather well.  However, it is the least
satisfactory model respective to the ratio of formation rates of
binary and single pulsars (see discussion section).

\begin{figure}
\centerline{
\psfig{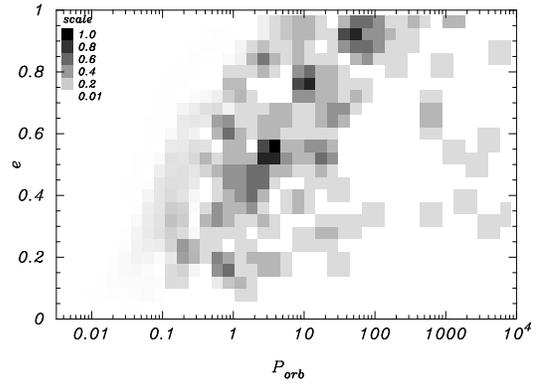}}
\caption{
The probability distribution for model D ($\elce = 0.5$, and 
kick according to Eq.~5) 
at the age of the population younger than 100\,Myr.
The gray scaling, given in the upper left corner, is chosen independently
of the other models and gives 8.9 (ns, ns) binaries in the Galaxy
for the darkest shade.}
\label{figeta1}
\end{figure}

Figure~\ref{figeta1} gives the \pe plane for the population 
younger than 100\,Myr for (\ns, \ns) systems of model~D. 
In this model the efficiency for the
deposition of orbital energy into the common envelope during the
spiral-in is considerably smaller than in the standard model: $\elce =
0.5$. The birthrate of (\ns, \ns) binaries in this model is considerably
smaller than in the standard model, as a result of greater proportion
of mergers under low $\eta_{\rm ce}.$~ However, all four HMBPs
appear
in low-probability areas of the diagram.  This was the basic reason for
pushing $\eta_{\rm ce}$ to higher values in our efforts to obtain
satisfactory models for the population of binary neutron stars.

\begin{figure}
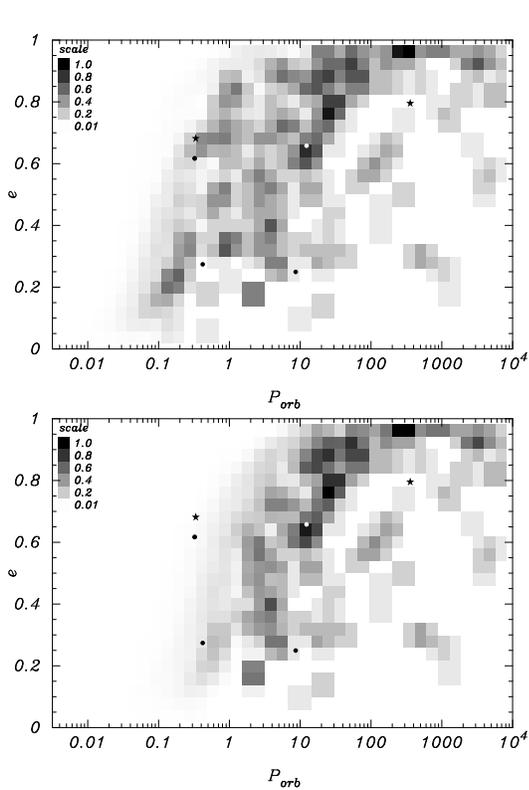
 
\centerline{
\psfig{file=GBa4q-1fpp_T2.ps,bbllx=575pt,bblly=40pt,bburx=95pt,bbury=705pt,height=5cm,angle=-90}}
\centerline{
\psfig{file=GBa4q-1fpp_T4.ps,bbllx=575pt,bblly=40pt,bburx=95pt,bbury=705pt,height=5cm,angle=-90}}
\caption{
The probability distribution  
for the Galactic number of (\ns, \ns) binaries younger than 100\,Myr 
(upper panel) and 10\,Gyr (lower panel) in model H.
The normalization gives 6.3 (\ns, \ns)  for the darkest shades in the 
upper panel and 522 for the lower panel}
\label{figGBrown}
\end{figure}

Figure~\ref{figGBrown} gives the \pe\ distribution of the neutron star
binaries which originate from model H for two age limits. 
This distribution is hardly affected by age 
(compare with model B in Fig.~\ref{figPPkick}).
Since systems which experience a common envelope are
absent in this model, the average orbital separation is relatively large.
The distribution is static for systems with $P_{orb} \ga 1$~day; 
it is hard to discriminate between the 100\,Myr
and the 10\,Gyr populations.

The population which originates from model~H is
a subset of model B, and Fig.~\ref{figGBrown} consequently
represents the \pe\ distribution of this subset in
Fig.~\ref{figPPkick}.
Only the young pulsar PSR~B2303+46 is present in the area with the highest
probability in Fig.~\ref{figGBrown}. Therefore
it is appealing to suggest that it originates from 
scenario~\II. 
Similarly, van den Heuvel et al (1994) argue that the evolution of the
progenitor of this system was dominated by mass loss in a stellar wind
rather than by Roche-lobe overflow.
A common envelope phase is required to form short 
period binary pulsars.
The apparent ``success'' of Wettig \& Brown (1996) and Fryer \&
Kalogera (1997) in reproducing PSR~B1913+16 and B1534+12 is mainly due
to their fine tuning of the parameters of the pre-supernova binary.
Such systems are not the dominant contributors to short orbital
period HMBP.

\begin{figure}
\centerline{
\psfig{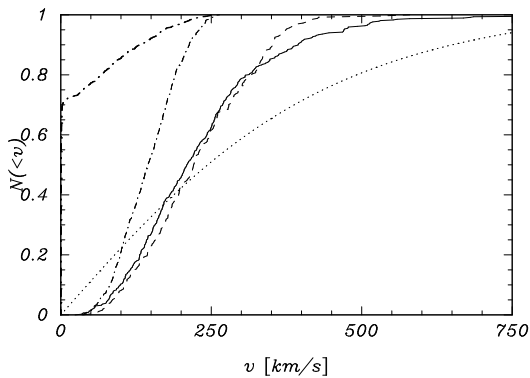}}
\caption{
The cumulative distribution of the center-of-mass velocities for neutron
star binaries which merge within 10~Gyr from models A (dash-dotted
line to the right), B (solid 
 line) and C (dashed line). 
For model A, for comparison, the line to the left represents
the velocity distribution of {\em all} (\ns, \ns) binaries and includes a
rather large population of systems  which do not merge within 10~Gyr.
The dotted line represents the input kick-velocity distribution for
model B (see Eq.~5).}
\label{figvcmdist}
\end{figure}

Figure~\ref{figvcmdist} gives the cumulative velocity distribution for
the center of mass of the merging (\ns, \ns) binaries which results from the
model computations. For comparison, we also give the kick distribution
(Eq.~\ref{eqkick}).  As expected, the model without the kick produces
low velocity binary neutron stars and models B and C produce the
higher velocity ones.  The low-velocity shoulder for the model without
a kick (left dash-dotted line), shown for comparison,
is the result of very wide
-- non-interacting -- binaries which happen to stay bound due to the
orbital eccentricity and low mass loss during the supernova.  The
short period (\ns, \ns) binaries from the model without a kick reach
much higher velocities (right dash-dotted line), but by far not as
high as in the models where a kick is imparted to the newly formed
neutron star.  The difference between the models with a Maxwellian
kick and a \P\ kick becomes quite noticeable in the tail of the
distribution: model B produces more high velocity neutron star
binaries.  The velocity distribution for HMBP which are produced in
model H is very similar to that of model B (solid line) but it is
shifted to lower velocities by about 50\,km/s, as a result of the
absence of very close systems among their progenitors.  The velocity
distribution for the single pulsars is similar to the kick
distribution, the contamination by recycled pulsars (which typically
have smaller velocities, Hartman \etal\ 1997b) is negligible.  The
distribution for HMBP has a lower proportion of high-velocity objects
compared to the single pulsars because the high velocity tail of the
kick distribution prevents the formation of (\ns, \ns) binaries.
There are not references to the spatial velocities of binary pulsars
in the refereed literature and it is impossible to confront
Fig.~\ref{figvcmdist} with observations. We should note, however, that
discovery of a HMBP with velocity in excess of $\sim 250\, \kms$~
would argue for the presence of kicks.

\section{Discussion}
\subsection{Populaton of HMBP and their merger times}

Population synthesis computations for the formation and further
evolution of binary neutron stars provide estimates for their birthrate
and the expected orbital elements of the present day population. 
The most crucial parameter for modeling of (\ns, \ns) binaries is the
common envelope parameter.
This parameter is not strongly constrained by observations.  
Population synthesis and smoothed particles hydrodynamic
computations suggest that energy
deposited into common envelopes may
be high compared to the orbital binding energy.
Our preferred model B is characterized by such a high efficiency:
$\elce = 2$. Higher values of $\elce$ up to 4 still
result in agreement with the observed population 
but the birthrate of neutron
star binaries becomes uncomfortably high compared to the formation
rate of single radio pulsars.
For $\elce \la 1$ 
the birthrate of neutron star binaries decreases significantly
but the period-eccentricity diagram fits badly with the four 
observed systems.
However, it might well be that \elce\ depends on the evolutionary
stage of the binary; for an inspiralling neutron star this parameter
might have a very different value than for a
main-sequence star or white dwarf. Neutron stars can be very efficient
in dispelling the common envelope by energy released by \eg\  
small explosions (Fryer \etal\ 1996) or semi-steady burning on 
its surface during the spiral-in process.

Only the models in which a kick is imparted to the newly born neutron
star from the distribution proposed by \P\ (1990) with parameters
suggested by Hartman \etal (1997a) give a satisfactory birthrate as
well as a distribution over $P_{orb} - e.$\  Increasing the fraction of
low-velocity kicks in \P\ distribution will however, increase the
quality of the fit of the \pe\ plane.  This distribution seems at the
moment preferable because the reproduction of transverse velocities of
young pulsars with measured proper motions requires the presence of a
significant number of objects with low velocities (Hansen \& Phinney 1996,
Hartman \etal 1997a).  The actual kick velocity distribution might
have an even larger contribution of low velocity kicks than
Eq.~\ref{eqkick} suggests.
 
Apart from kick velocity, all other parameters have
a relatively small effect. Birth- and merger rates vary within a
factor of a few, but not by orders of magnitude. Therefore it is
hard to decide on population estimates of birthrates which model
parameters are favoured.

The observations may be best fit with the computed probability
distribution in the \pe\ plane if the age of the recycled neutron
stars is of the order of 100 million to 1 billion year.  Note that
there might well be effects other than the age alone which have its
influence on the detectability of neutron star binaries; the pulse
period, the strength of the magnetic field and the luminosity.  All
these parameters give rise to a different lifetime of the pulsar and
different observational characteristics which make the selection
effects hard to quantify.  It is appealing that the youngest system
PSR~B2303+46 fits the relatively high probability region of \pe\
diagram for the youngest pulsars (Fig.~\ref{figzero_age}, lower
panel), the oldest one PSR~J1518+49 fits best if the age of the
population is as old as a few billion years (close to the one in the
lower panel of Fig. ~\ref{figPPkick}) and intermediate age pulsars
PSR~B1913+16 and PSR~B1534+12 are best reproduced by a population of a
few hundred Myr (Fig.~ \ref{figPPkick}).  Since the best fit between
the observations and the models is at an age of the population between
100~Myr and 1~Gyr, this suggests that the majority of the recycled
neutron stars are born with low magnetic field.

Taking into consideration 
the characteristic age, orbital period and the
eccentricity of PSR~B2127+11C, one may suggest that it is formed by
scenario \I instead of an exchange interaction in the high-density
core of the globular cluster M15 (see Phinney \& Sigurdsson 1991).  
However, to be
primordial, the pulsar should be at least 10\,Gyr old 
(the age of the globular cluster).  This pulsar
has to merge with its companion within a few 100\,Myr and it is highly
unlikely to observe it in the last percent of its lifetime as a 
(\ns, \ns) binary.

The birthrate of binary pulsars relative to singles is $\sim
1/430$ in the standard model, while in the most extreme model H it is
$\sim 1/1775$. These numbers have to be reduced for the binary
fraction, which is smaller than 100\%.  According to \cite*{vdhl96}
the lifetime of recycled pulsars is a factor $\sim 3$ larger than
their characteristic age and therefore it is between $\sim 100$~ Myr
and $\sim 1$~ Gyr.  On the other hand population synthesis studies
suggest that $\sim 90\%$ of the recycled pulsars are missed in
pulsar surveys (\cite{cl95}).  The lifetime of young pulsars is $\sim
10$\,Myr.  Combining these numbers we arrive at the conclusion that
the relative birthrate of single pulsars and binary pulsars also
reflects their relative galactic number density.  If beaming factors
are similar for old and young pulsars, relative numbers of \ns\ and
(\ns, \ns) in our models do not contradict observations (see Bailes
1996).  As the birthrate of single pulsars in our model is consistent
with the rate of supernovae and the birthrate of single pulsars, we
may conclude that our models give basically the actual Galactic birthrate
of binary neutron stars.

A fraction of the neutron stars in binaries may never become
observable as recycled pulsars.
The large pulse periods of X-ray pulsars suggest that observed recycled
pulsars are resurrected at later stages. 
It may be shown that in the common-envelope stage that follows the
X-ray pulsar stage, if $B \apgt 10^{11}$\, G, accretion at the Eddington
rate doesn't spin neutron stars to $P \aplt 100$\, ms (Ergma \&
Yungelson 1977).
Recycling of a pulsar to $P_{\rm eq} \sim 30$ ms requires the accretion of 
\mbox {$\Delta M \approx 0.17 \left(P_{\rm eq} / {\rm ms} \right)^{-4/3} 
\approx 0.002$\msun} (\eg\ Bhattacharya 1996). 
The  only evolutionary stage in which this can
happen is during accretion from the wind of the accompanying helium star.
The ``propeller mechanism'' (Illarionov \& Sunyaev 1975), 
which prevents accretion onto the neutron star may
cause the neutron star to remain dead.
Ergma \& Yungelson (1997) derive an estimate for
the orbital period at which the propeller mechanism is effective:
\begin{equation}
P_{\rm orb} \apgt 0.5 M_{\rm ns}^{3/2} v_{1000}^{-3} (M_{\rm ns} +
M_{\rm he})^{-1/2} M_{\rm he}^{15/8}~~{\rm hr},
\end{equation}
where $v_{1000}$ is the velocity of the helium-star wind in 1000\,\kms\ 
and masses are in~ \msun.
The spin slow-down time in the (\ns, \he) systems 
is comparable to the helium burning time. 
For typical $v_{1000} \sim 1$ -- 2, the critical period is 
of the order of only several hours for a low mass helium star, 
which is quite short. 
If the propeller mechanism is effective, accretion is possible for 
short period binaries with a low mass helium star or for wide systems
with a high mass helium star.
In the computations of Ergma \& Yungelson it appeared that the
majority of the neutron stars are not able to accrete from
helium companion winds
\footnote{The results of Ergma \&
Yungelson cannot be directly applied
here for numerical estimates, since their population synthesis was 
made under different assumptions. 
Qualitatively it is clear that for the present
computations accretion would be prohibited in even larger proportion of
all systems, since here we typically produce wider (\ns, \he)
binaries.}.
The first born pulsar has plenty of time (10 to 20~Myr) 
to die before its companion becomes a neutron star.
If the propeller mechanism is effective, the first born pulsar will
never be resurrected from the graveyard.
In this case the birthrate of recycled pulsars in binaries may be
smaller than the birthrate of neutron star binaries.
According to Lipunov \etal\ (1997) this fraction may be as small as 
$\sim 1$\%, depending on the assumed initial magnetic moment of the
neutron star and the decay timescale of its magnetic field.
There may be one such a binary among the observed pulsars, and in
fact it is possible that PSR~B2303+46 contains  a non-revived 
pulsar next to a young one (see Table~\ref{tabobserved}).

An observational estimate for the birthrate of HMBP can be derived
from the estimate of their number in the Galaxy and their lifetimes as
pulsars (Kulkarni \& Narayan 1988).  Using scaling factors,
correction factors for incompleteness of the sample and beaming factor
from Curran \& Lorimer (1995) and lifetimes from 
van den Heuvel \& Lorimer (1996) for
PSR~B1534+12, B1913+16, and B2303+46 we arrive at a birthrate of 
$2.7 \times 10^{-5}$\,\pyr. This is comfortably close to our standard
model B (Table~\ref{tabbr}).  This derived birthrate heavily relies on
the young PSR~B2303+46.  Excluding it reduces the birthrate to
$0.8 \times 10^{-5}\,\pyr$, which is obviously the estimate obtained by
van den Heuvel \& Lorimer (1996).  There is no data available on
scaling factor for PSR~J1518+49, but we may expect that it will enter
the estimate of the birthrate with a very small weight because of its
long characteristic age.

\begin{figure}
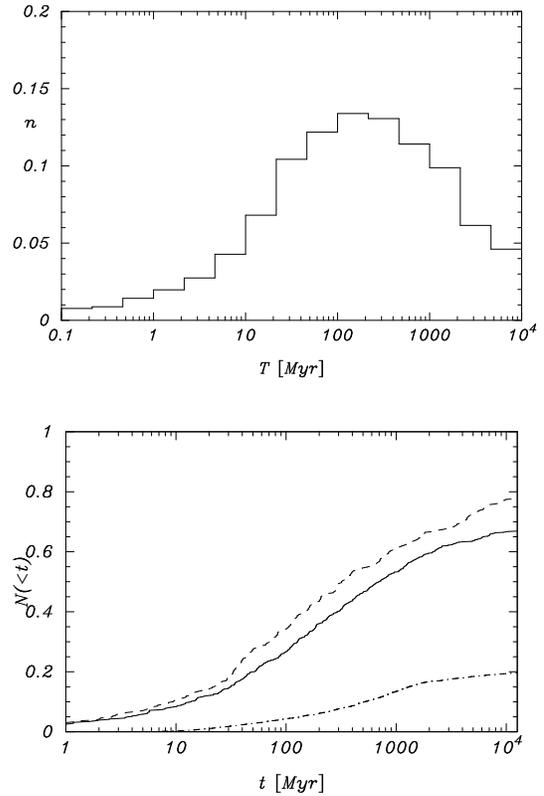

\vspace{-0.5cm}
\centerline{
\psfig{file=nsmBage_hist.ps,bbllx=575pt,bblly=40pt,bburx=95pt,bbury=705pt,height=5cm,angle=-90}}
\vspace{0.5 cm}
\centerline{ 
\psfig{file=nsnsTmerge_cum.ps,bbllx=575pt,bblly=40pt,bburx=95pt,bbury=705pt,height=5cm,angle=-90}}
\caption[]{
{\it Upper panel} - Normalized distribution of (\ns, \ns) binaries over
merger ages in model B. The limit of 10 Gyr is adopted as the age of
the galactic disc. {\it Lower panel} -  The cumulative distribution of the (\ns, \ns) binaries over merger
ages for the models A (dash dotted line),~B (solid line) and C (dashed
line). 
The fraction of mergers reached within $10^{10}$~yrs indicates that a
considerable fraction of systems does not merge within a Hubble time.
}
\label{figage}
\end{figure}

To summarize, we may argue that our standard model B, which reproduces
best the orbital periods and eccentricities of the observed HMBP is
not in conflict with observations also in respect to the birthrate
of these systems.  Therefore we may expect, that this model gives a
realistic estimate of the merger rate of neutron star binaries 
$\approx 2 \times 10^{-5}$\,\pyr.  This number is not inconsistent with
the estimated rate of $0.8 \times 10^{-5}$\,\pyr\ from van den
Heuvel \& Lorimer (1996) based on two observed pulsars with merger
times shorter than the Hubble time and the assumption of a steady
state in which the birthrate equals the merger rate.  Actually, from
Table~\ref{tabbr} and Figs.~1 and 7 it is immediately clear that there
does not exist a steady state in which the birthrate equals the merger
rate.
About 10\% to 20\% of all (\ns, \ns) binaries merge after the recycled
pulsar has died within 0.5 -- 1\,Gyr (see Fig~\ref{figage}). 
These binaries hang around undetected in the
Galaxy until they merge.  
A considerable fraction of these (20\% to 30\%) are parked in wide
orbits in the Galactic halo or escape from its potential well (see
Fig.~\ref{figDmax} and \ref{figZgal}). 
Another $\sim 10\%$~ merge within $\sim 10$ Myr, before even the young
pulsar dies (see Fig~\ref{figage}).  These systems may also avoid
detection due to 
their short lifetime (their number relative to the number of single pulsars
is only $\sim 1/4000$~ in model B for 100\% binarity).  
Figure~\ref{figage} suggests a
revision of the ``observational'' estimate upward by a factor 
$\sim 1.5$ to $1.2 \times 10^{-5}$.  
This correction factor my increase even further taking 
non-recycled, hence non-observable, close binary neutron stars into account.
(Curiously enough the situation with merging neutron stars is
similar to that of merging white dwarfs which are candidate
progenitors for SNe~Ia; most of the mergers will occur when both white
dwarfs have cooled beyond detectability, Iben \etal 1997.)
    
On the other hand, about 1/3 of all (\ns, \ns) binaries born annually
in the Galaxy have merger times exceeding 10\, Gyr and are
accumulating in it.  The total number of dead non-merging pairs
present in the Galactic disc may be as large as $10^5$.

Figure~\ref{figage} indicates that the Galactic disc merger rate is
determined mainly by the star formation rate during the last
1--2\,Gyr.  The average star formation rate over the history of the
Milky Way is about 4 times higher than the current rate (\eg\ van den
Hoek 1997). The same is the case for other massive ($ \sim 10^{11}$\,
\msun) spiral and irregular galaxies (Gallagher \etal 1984; Sandage
1986).  The star formation rate was much higher than present only in
the first few Gyr of the existence of the Galaxy.  Figure~\ref{figage}
shows that this early star burst contributes only little to the
current merger rate of (\ns, \ns) binaries.  The same point has to be
made with respect to elliptical galaxies. As most stars in them were
formed in a short ($\aplt 1$\,Gyr) burst, overwhelming majority of
merger candidates already merged long ago.  Figure~\ref{figage}
suggests that in a galaxy which was formed in a 1\,Gyr burst, the
present rate of (\ns, \ns) mergers per unit mass is $\sim (10-15)\%$
of that in the galaxy of the same mass but formed in 10\,Gyr.  In the
extrapolation of the Galactic rate of merger to the local Universe the
different contributions from spirals and ellipticals have to be
considered (see Bagot 1996 for a parametric study of this issue).
Scaling to the total number of galaxies (spirals as well as
ellipticals) within 100\,Mpc after Curran \& Lorimer (1995) gives an
upper limit for the number of mergers detectable by the advanced LIGO
detector of 1\,\pyr\ for model B.

\subsection{Merger rate in the Universe and $\gamma$-ray bursts} 

A rough estimate of the total number of merger events in the Universe
may be obtained from
scaling the Galactic merger rate by the ratio of blue-band luminosities of
the Milky Way and the Universe: $R \approx 10^{-2} h\,{\rm
Mpc^{-3}},$~ where $h$ is Hubble constant in 100 \kms (Phinney
1991). For $H_o = 75\,\kms {\rm Mpc^{-1}}$~ this gives, extrapolating
model B, $\sim 4.6 \times 10^4$ events per year. This number may be
compared to the number of $\gamma$-ray bursts, which are also
suggested to originate from neutron star mergers (\P\ 1986). The
observational estimate of the rate of detectable $\gamma$-ray
bursts in the Universe is several times $10^{-6}$~ per year per galaxy
(Mao \& \P\ 1992; Piran 1996).  The discrepancy with the observed
rate (for model B) may be erased if $\gamma$-ray emission is beamed
into an opening angle of $\sim 10^{\circ}$ (Mao \& Yi
1994; Piran 1996).  The anisotropy of the energy release in
$\gamma$-ray production is suggested by numerical merger models
(Davies \etal\, 1994).

\begin{figure}
\centerline{
\psfig{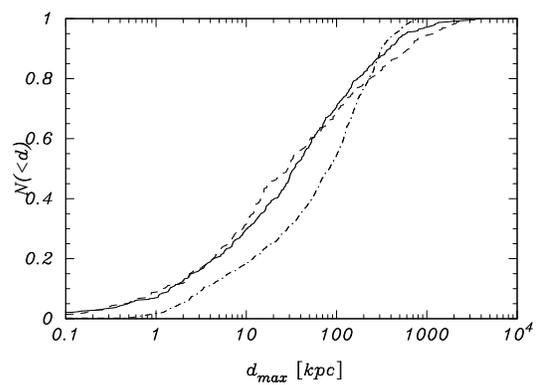}}
\caption[]{
The cumulative distribution of (\ns, \ns) binaries over the distance
traveled before coalescence occurs.
The line styles represent the results from the same model computations
as denoted in Fig.~\ref{figage}. 
}
\label{figDmax}
\end{figure}

\begin{figure}
\centerline{
\psfig{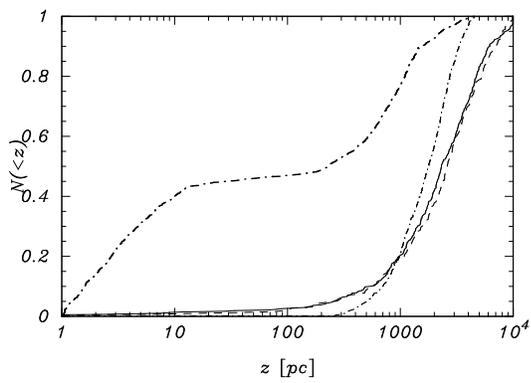}}
\caption[]{ The cumulative distribution of (\ns, \ns) binaries over
the galactic height $z$ reached before coalescence occurs.  The
galactocentric distance at which the merger occurs is computed using
Eq.~\ref{zv_relation}.  The line styles are as in
Fig.~\ref{figvcmdist}.  The subset of (\ns, \ns) binaries from model A
which merge within 10~Gyr (right dash-dotted line) reach higher
$z$-values then those with orbital period so large that they do not
merge within the age of the Universe (left dash-dotted line).  }
\label{figZgal}
\end{figure}

Virtually all (\ns, \ns) binaries in the models with a kick escape
from the potential well of a globular star cluster (see
Fig.~\ref{figvcmdist}) and a considerable fraction is expected to
escape from a dwarf galaxy.  In the model without a kick, this
fraction is only $\sim 25\%$ (about $10\%$\ for dwarf galaxies). But
in the latter model the majority of (\ns, \ns) systems originate from
wide non-interacting binaries which never produce a recycled pulsar
and which will not merge within the age of the Universe. The (\ns,
\ns) pairs confined to parental stellar systems have typically large
orbital periods and are therefore very fragile for three-body
encounters.  The result is that globular clusters 
are deficient of HMBP's according to the models
with a kick and also according to the models without a kick!
In the presence of a
kick the formation of (\ns, \ns) binaries is not an efficient way to
keep a neutron star bound to a globular cluster.

The maximum distance traveled by a (\ns, \ns) binary before it merges
may be computed using the distribution functions for the merger time
and the center-of-mass velocity of the binaries. Figure~\ref{figDmax}
provides this distribution for the HMBPs.  It is interesting to notice
that in the models without a kick the typical distance traveled is
considerably larger than in the models with a kick. Note also that
the distributions for both models with a kick are indistinguishable.
The binaries in the high-velocity tail of model B (see
Fig.~\ref{figvcmdist}) merge on a short time scale (see
Fig.~\ref{figage}) whereas the low-velocity binaries in the model with a
Maxwellian kick live rather long.

The distance traveled perpendicular to the galactic plane 
in the potential of the Galaxy can be estimated from a relation
between the $z$-component of
the initial velocity $v(z)$ of an object and the 
maximum galactic height $z$ 
(Tutukov \& Yungelson 1978):  
\begin{equation}
v(z) \approx \left[ 6 z - 2000 \ln (1+0.003 z) \right]^{1/2},
\label{zv_relation}\end{equation}
where $v$\ is  in \kms\ and $z$\ in pc.   
The maximum $z$ reached by a (\ns, \ns) binary before coalescence
was estimated using the velocity from the distribution given in 
Fig.~\ref{figvcmdist} divided by $\sqrt{3}$ (assuming that the
velocity is distributed isotropically).
To correct for the pre-merger lifetimes which may be shorter 
than the travel time to the maximum $z$,
the minimum of $z$ resulting from Eq.~\ref{zv_relation} and the
maximum distance traveled (Fig.~\ref{figDmax}) was adopted.

Figure.~\ref{figZgal} shows that for the models without a kick the
majority of the systems do not reach a high $z$ (left dash-dotted
line), as expected from the enormous low-velocity tail in
Fig.~\ref{figvcmdist}. The short period (\ns, \ns) binaries from model
A reach considerably higher Galactic heights (right dash-dotted line)
but by far not as high as in the models with a kick.  Note that the
majority of the binaries in the low-$z$ shoulder (left dash-dotted
line in Fig.~\ref{figZgal}) are not visible as systems with a recycled
pulsar, since they did not experience a phase of mass transfer.  The
(\ns, \ns) binaries from the models where a kick is imparted to the
newly formed neutron star show a much more extended $z$
distribution. Actually, a considerable fraction of these systems
escape from the potential of the Galaxy or are parked in the Galactic
halo.  Escape of some (\ns, \ns) beyond the sensitivity limits of
detectors may be a factor contributing to the underestimate of the
galactic birthrate of HMBP.  Since the life times of single radio
pulsars are considerably shorter than that of recycled pulsars, of
which the lifetime is comparable to the merger time, the galactic
heights of the latter are expected to be considerably larger.

Even in the absence of kicks (\ns, \ns) binaries, which escape mainly
from dwarf galaxies, may form a common intergalactic field of these
objects (Tutukov \& Yungelson 1994). It is suggested that most of star
formation in the Universe occurs at low redshift ($z \sim 1$, see
e.g.\ Madau 1997, Connolly \etal\ 1997) and the main sites of star
formation could be dwarf star-burst galaxies (\eg\ Babul
\& Ferguson 1996).  However, due to the large distances traveled by
the (\ns, \ns) before they merge, a significant fraction merge at a
large distance form the host galaxy in which the binary was formed.
If $\gamma$-ray bursts are related to (\ns, \ns) mergers, a
considerable fraction may occur at large distance from their host
galaxy and they are not likely to be associated.

\subsection{Black hole plus neutron star binaries}

Our models suggest (Table~\ref{tabbr}) that the number of detected
radio pulsars has to be at least 3 to 4 times larger than at present
before one may hope to find a young pulsar accompanied by a black
hole. The only exception is in model H in the version in which
hypercritical accretion onto the neutron star does not always result
in swallowing of the companion. In this model one may expect the
presence of 2 or 3 (\bh, \ns) systems among the known young pulsars
(see Table~\ref{tabbr}).  The number of (\bh, \ns) binaries has then
to be comparable to the number of HMBPs.  Both these conclusions are
in conflict with the observations.  The merger rate of (\bh, \ns)
binaries is typically more than an order of magnitude smaller than
that of (\ns, \ns) binaries.  An exception to this is model D with
small \elce, which prevents the formation of (\ns, \ns) systems by
merging their progenitors in the common envelope phase.

For gravitational waves detectors the signal-to-noise ratio is
(Bonazzola \& Marck 1994)
\begin{equation} 
S/N \propto \frac{(M_1M_2)^{1/2}} {(M_1+M_2)^{1/6}}D^{-1},
\label{sneq}
\end{equation}
where $M_{1,2}$\ are the masses of coalescing objects and $D$\ is the
distance to them. E.g., for the range of black hole masses 4 to 17
\msun\ suggested by black-hole candidate binaries, one would expect
that for (\bh, \ns) mergers detection distance may be 1.5 to 2.5
larger than for (\ns, \ns) mergers. Hence, the rates of detectable
(\ns, \ns) and (\bh, \ns) mergers are expected to be comparable (about
1/yr). This has to be 
considered in the context of expected coalescence waveforms from the
first detections (Thorne 1996). Also, the characteristics of
$\gamma$-ray bursts expected from (\bh, \ns) mergers may differ from
the ones from (\ns, \ns) mergers (\P\ 1991).

Due to the large orbital period of (bh, bh) binaries, their expected merger 
rate is negligible and this number is not given in Table~\ref{tabbr}.
A rate of detectable (\bh, \bh) mergers comparable to the rate of (\ns,
\ns) mergers was predicted by \eg\ Tutukov \& Yungelson (1993a) 
and Lipunov \etal (1997).
According to our current understanding this is
a result of underestimating the effect of 
high mass loss in the stellar winds of stars more massive than 
$\sim 40$\,\msun.

\section{Conclusions}

We discussed in detail the formation and evolution of binaries which
contain two neutron stars or black holes.  The orbital parameters of
present-day population in the Galaxy are studied.  For a successful
model we require that also the results of the preceding evolutionary
stages (occurrence rates for supernovae and the formation of single
pulsars, number of Be and high-mass X-ray binaries) comply with
observations.  As an additional restriction we require the observed
high-mass binary pulsars to fit into the modeled \pe\ diagram for the
appropriate age of the population.  We summarize our conclusions:

\begin{itemize}
\item[1.]  
Deposition of orbital energy into common envelopes has to
be very efficient for in-spiraling neutron stars, and sources other
than orbital energy have to be invoked.

\item[2.]
A kick velocity distribution with a high velocity tail and the bulk
in small velocities (as given by \P\,1990 with the
parameters from Hansen \& Phinney 1996) gives a model which is in
agreement with the observations. The same velocity distribution provides a
satisfactory model for single pulsars (Hartman 1997).
Increasing the fraction of low velocities may provide an even better
fit to the observations.

\item[3.]
The best fit between the observations and the models is at an
age of the population between several 100\,Myr and 1~Gyr.
This suggests that
the majority of the recycled neutron stars are born with low magnetic
fields. 

\item[4.]
In the model of the population of (\ns, \ns) binaries which we
consider satisfactory, their birthrate is $\sim 3.4 \times
10^{-5}$\,\pyr\ (assuming 100\% binarity).  This is consistent with the
birthrate derived from observed binary pulsars.  The merger rate of
(\ns, \ns) binaries in this model is $\sim 2 \times
10^{-5}$\,\pyr. The merger rate of (\bh, \ns) binaries is smaller
by at least an order of magnitude, but they may contribute significantly
to the rate of expected gravitational wave detections because of the higher
total mass of the systems.

\item[5.]
Binary neutron stars merge typically 
within 1 to 2\,Gyr after formation. Hence, elliptical 
galaxies and the old stellar population in spiral and irregular galaxies 
may not contribute significantly to the rate of 
gravitational waves detections.
When the detectors become sensitive to the events at
cosmological distances this population may give an important contribution.

\item[6.]
The rate of (\ns, \ns) mergers is consistent with the estimated
rate of $\gamma$-ray bursts if the latter are beamed into an opening
angle of $\sim 10^{\circ}$.

\item[7.]
The scenario for the evolution of close binary stars which assumes 
hypercritical accretion and the subsequent collapse of the neutron star into a
black hole in the common envelope is able to reproduce
binary neutron stars with large orbital periods ($\sim 10$\,days), but 
fails to reproduce the short period systems.

\item[8.]
In the scenario in which a neutron star accretes hypercritically
and the binary avoids merging,
the Galactic population of binaries which
contain a black hole and a neutron star might well exceed
the number of double neutron stars. The merger rate of these
binaries is then approximately  $3 \times 10^{-5}\,\pyr.$ 

\item[9.]
The number of observed single radio pulsars has to increase to
about 2000, before one can hope to find a young radio pulsar
accompanied by a black hole.

\item[10.]
A considerable fraction of the binary neutron stars escape from the
tidal field of the Galaxy and travel up to $\sim$\,Mpc
distances before they coalesce.

\item[11.]
Short period neutron star binaries are, due to the high
velocities received upon the supernova, virtually all ejected from the 
shallow potential well of globular star clusters. 
High-mass binary pulsars in globular clusters are therefore likely to
be formed by a dynamical interaction.

\end{itemize}

\acknowledgements 
We thank Dipankar Bhattacharya, Gerry Brown, 
Ene Ergma, Hanno Spreeuw, Alexander
Tutukov, Ed van den Heuvel
and Frank Verbunt for numerous stimulating discussions.  Ed
van den Heuvel is acknowledged for critically reading the manuscript.
SPZ acknowledges Ramesh Narayan for the suggestion to use the entire
\pe\ plane for a parametric study of the (\ns, \ns) population.  LRY
acknowledges support through the NWO Spinoza grant and the warm
hospitality of the Astronomical Institute ``Anton Pannekoek'' and
Meudon observatory. This
work was supported by NWO Spinoza grant 08-0 to E.~P.~J.~van den Heuvel
and RFBR grant 960216351.


\begin{thebibliography}{}
 
\bibitem[\protect\astroncite{Abt}{1983}]{a83}
Abt, H. A., 1983, ARA\&A 21, 343

\bibitem[\protect\astroncite{Babul \& FergusonF}{1996}]{bf96}
Babul, A., Ferguson, H. C., 1996, ApJ, 458, 100
\bibitem[\protect\astroncite{Bagot}{1996}]{bag96}
Bagot, P., 1996, Ph.\ D.\ Thesis, University of Montpellier II.

\bibitem[\protect\astroncite{Bagot}{1997}]{bag97}
Bagot, P., 1997, A\&A, 322, 533

\bibitem[\protect\astroncite{Bailes}{1996}]{ba96}
Bailes, M., 1996, in Compact stars in binaries, ed. J. van Paradijs, E. 
P. J. van den Heuvel, E. Kuulkers (Dordrecht: Kluwer), 
p. 213

\bibitem[\protect\astroncite{Bhattacharya}{1996}]{bh96}
Bhattacharya, D., 1996, in Compact stars in binaries, 
eds. J. van Paradijs, 
E. P. J. van den Heuvel, E. Kuulkers (Dordrecht: Kluwer), p. 243

\bibitem[\protect\astroncite{Bhattacharya \& van~den Heuvel}{1991}]{bh91b}
Bhattacharya, D., van~den Heuvel, E., 1991, PhysRep, 203, 1

\bibitem[\protect\astroncite{Bisnovatyi-Kogan \& Komberg}{1975}]{bkk74}
Bisnovatyi-Kogan, G. S., Komberg, B. V., 1975, SvA, 18, 217

\bibitem[\protect\astroncite{Bonazzola \& Marck}{1994}]{bz94}
Bonazzola, S., Marck, J.-A., 1994, Ann. Rev. Nucl. Part. Sci., 45,
655
  
\bibitem[\protect\astroncite{Brandt \& Podsiadlowski}{1995}]{bp95}
Brandt, N., Podsiadlowski, P., 1995, MNRAS 247, 484
  
\bibitem[\protect\astroncite{Brown}{1995}]{bro95}
Brown, G., 1995, ApJ, 440, 270

\bibitem[\protect\astroncite{Brown, Weingartner \& Wijers}{1996}]{bww96}
Brown, G., Weingartner, J. C., Wijers, R. A. M. J. 1996, ApJ, 463, 297


\bibitem[\protect\astroncite{Clark \& Eardley}{1977}]{clae77}
Clark, J. P. A., Eardley, B., 1977, ApJ, 215, 311 



\bibitem[\protect\astroncite{Clark etal}{1979}]{chs79}
Clark, J. P. A., van den Heuvel, E. P. J., Sutantio, W., 1979, A\&A,
72, 120

\bibitem[\protect\astroncite{Cappellaro \etal}{1997}]{cap97}
Cappellaro, E. Turatto, M., Tsvetkov, D. Yu., \etal 1997, A\&A, 322, 431

\bibitem[\protect\astroncite{Carraro \& Chiiosi}{1994}]{cc94}
Carraro, G., Chiosi, G., 1994, A\&A, 287, 761

\bibitem[\protect\astroncite{Chevalier}{1993}]{ch93}
Chevalier, R. 1993, ApJ, 411, L33

\bibitem[\protect\astrncite{Connolly \etal}{1997}{c97}]{}
Connolly A.J. \etal\, 1997, ApJ 486, L11

\bibitem[\protect\astroncite{Curran \& Lorimer}{1995}]{cl95}
Curran, S. J., Lorimer, D. R., 1995, MNRAS, 276, 347

\bibitem[\protect\astroncite{Davies \etal}{1994}]{dav94}
Davies, M.\ B., Benz, W. Piran, T., Thielemann, F.\ K., 1994, 
ApJ, 431, 742

\bibitem[\protect\astroncite{Deich \& Kulkarni}{1996}]{deku94}
Deich, W. T. S., Kulkarni, S. R., 1996, 
in Compact stars in binaries, ed. J. van Paradijs, E. P. J. van den 
Heuvel, E. Kuulkers (Dordrecht: Kluwer),  p. 279

\bibitem[\protect\astroncite{De Loore}{1975}]{del75}
De Loore, C.W.E., De Greve, J.P., van den Heuvel, E.P.J., De Cuyper,
J.P., 1975, Mem. Soc. Astron. It., 45, 893

\bibitem[\protect\astroncite{Deshpande \etal}{1995}]{drg95}
Deshpande, A. A., Ramachandran, R., Srinivasan, G., 1995, JA\&A, 16, 53 

\bibitem[\protect\astroncite{Dewey \& Cordes}{1987}]{dc87}
Dewey, R. J., Cordes, J. M., 1987, ApJ, 321, 780


\bibitem[\protect\astroncite{Duquennoy \& Mayor}{1991}]{dm91}
Duquennoy, A., Mayor, M., 1991, A\&A, 248, 485

\bibitem[\protect\astroncite{Ergma \& van den Heuvel}{1997}]{evh97}
Ergma, E.,  van den Heuvel, E. P. J., 1997, A\&A, submitted

\bibitem[\protect\astroncite{Ergma \& Yungelson}{1997}]{ey97}
Ergma, E.,  Yungelson, L. R.,  1997, A\&A,  in press

\bibitem[\protect\astroncite{Flannery \& 
van den Heuvel}{1975}]{fvdh75}
Flannery, B. P., van den Heuvel. E. P. J., 1975, A\&A, 39, 61
  
\bibitem[\protect\astroncite{Fryer \& Kalogera}{1997}]{fk97}
Fryer, C. L., Kalogera, V., 1997, ApJ, 489, 244

\bibitem[\protect\astroncite{Fryer \etal}{1996}]{fbh96}
Fryer, C. L., Benz, W., Herant, M., 1996, ApJ, 460, 801

\bibitem[\protect\astroncite{Gallagher \etal.}{1984}]{ght84}
Gallagher, J., Hunter, D., Tutukov, A. V., 1984, ApJ, 284, 544

\bibitem[\protect\astroncite{Gunn \& Ostriker}{1970}]{gu70}
Gunn, J., Ostriker, J., 1970, ApJ, 160, 979

\bibitem[\protect\astroncite{Habets}{1985}]{hab85}
Habets, G., 1985, PhD Thesis, U. Amsterdam
 
\bibitem[\protect\astroncite{Hartman}{1997}]{har97}
Hansen, B. M. S., Phinney, E. S., 1996,  BAAS, 189, 7402

\bibitem[]{}
Hansen, B. M. S., Phinney, E. S., 1997, astro-ph/9708071

\bibitem[\protect\astroncite{Hartman et~al.}{1997a}]{hbwv97}
Hartman, J. W., Bhattacharya, D., 
Wijers, R. A. M. J., Verbunt, F.,1997a, A\&A, 322, 477

\bibitem[]{}
Hartman, J. W., 1997, A\&A, 322, 127

\bibitem[\protect\astroncite{Hartman et~al.}{1997b}]{hpzv97}
Hartman, J.~W., Portegies~Zwart, S.~F., Verbunt, F., 1997b, A\&A, 322, 477

\bibitem[\protect\astroncite{Helling \& De Loore}{1986}]{hd86}
Hellings, P., De Loore, C., 1986, in Evolution in galactic X-ray binaries, 
ed. J. Truemper \etal (Dordrecht: Reidel), p. 51  

\bibitem[\protect\astroncite{Hulse \& Taylor}{1975}]{ht75}
Hulse, R.~A., Taylor, J.~H., 1975, ApJ, 195, L51

\bibitem[\protect\astroncite{Iben \& Livio}{1993}]{il93}
Iben, I. Jr., Livio M., 1993, PASP, 105, 1373


\bibitem[\protect\astroncite{Iben \& Tutukov}{1996}]{it96}
Iben, I. Jr., Tutukov, A. V., 1996, ApJ, 456, 738 

\bibitem[\protect\astroncite{Iben \etal }{1995}]{ity95}
Iben, I. Jr., Tutukov, A. V., Yungelson, L. R., 1995, ApJSS, 100, 217

\bibitem[\protect\astroncite{Iben \etal }{1997}]{ity97}
Iben, I. Jr., Tutukov, A. V., Yungelson, L. R., 1997, ApJ, 475, 291

\bibitem[]{}
Illarionov, A. F., Sunyaev, R. A., 1975, A\&A, 39, 185

\bibitem[\protect\astroncite{Kalogera \& Webbink}{1996}]{kw96}
Kalogera, V., Webbink, R. F., 1996, ApJ, 458, 301

\bibitem[\protect\astroncite{Kornilov \& Lipunov}{1984}]{kl84a}
Kornilov, V. G, Lipunov, V. M., 1984a, SvA, 27, 167

\bibitem[\protect\astroncite{Kornilov \& Lipunov}{1984}]{kl84b}
Kornilov, V. G, Lipunov, V. M., 1984b, SvA, 27, 334

\bibitem[\protect\astroncite{Kulkarni, \& Narayan, R.}{1988}]{kn88}
Kulkarni, S., Narayan, R., 1988, ApJ, 335, 755

\bibitem[\protect\astroncite{Lamb \& Lamb}{1976}]{ll76}
Lamb, D.Q., Lamb, F. K., 1976, ApJ, 204, 168

\bibitem[\protect\astroncite{Langer }{1989}]{la89}
Langer, N., 1989, A\&A, 220, 135

\bibitem[\protect\astroncite{Latham \etal}{1984}]{lat84}
Latham, D. W., Schechter, P., Tonry, J.,  Bahcall, J. N.,
Soneira, R. M., 1984, ApJ, 281, L41

\bibitem[\protect\astroncite{Lipunov \etal}{1996}]{lpp96}
Lipunov, V. M., Postnov, K. A., Prokhorov, M. E., 1996, A\&A, 310, 489


\bibitem[\protect\astroncite{Lipunov \etal}{1997}]{lpp97}
Lipunov, V. M., Postnov, K. A., Prokhorov, M. E., 1997, MNRAS, 288, 245


\bibitem[\protect\astroncite{Livio}{1996}]{li96}
Livio, M., 1996, in Evolutionary processes in binary stars, ed. R. A. M. 
J. Wijers, M. B. Davies, C. A. Tout (Dordrecht: Kluwer), p. 141

\bibitem[\protect\astroncite{Lyne\& McKenna}{1989}]{lm89} 
Lyne, A.~G., McKenna, J., 1989, Nat, 340, 367  

\bibitem[\protect\astroncite{Madau}{1997}]{m97}
Madau, P., 1997, in The Hubble Deep Field, ed.\ M.~Livio, M.~Fall, 
P.~Madau (Cambridge: CUP), in press

\bibitem[\protect\astroncite{Mao \& Paczy\'nski}{1992}]{maop92}
Mao, S., \P, B., 1992, ApJ, 388, L45

\bibitem[\protect\astroncite{Mao \& Yi, I.}{1994}]{mayi94}
Mao, S., Yi, I.,  1994, ApJ, 424, L131

\bibitem[\protect\astroncite{Massevich \etal}{1979}]{mpty79}
Massevich, A. G., Popova, E. I., Tutukov, A. V., Yungelson, L. R.,
1979, Ap\&SS, 62, 451

\bibitem[\protect\astroncite{Meurs \& van den Heuvel}{1989}]{mvdh89}
Meurs, E. J. A., van den Heuvel, E. P. J., 1989, A\&A, 226, 88 

\bibitem[\protect\astroncite{Meynet \etal.}{1993}]{mmm93}
Meynet, G., Mermilliod, J.-C., Maeder, A., 1993, A\&AS, 98,477 


\bibitem[\protect\astroncite{{Nice} et~al.}{1996}]{nst96}
Nice, D.~J., Sayer, R.~W., Taylor, J.~H., 1996, ApJ, 466, L87

\bibitem[\protect\astroncite{Paczy\'nski}{1971}]{pa71}
Paczy\'nski, B., 1971 Acta Astr., 21, 1

\bibitem[\protect\astroncite{Paczy\'nski}{1986}]{pa86}
Paczy\'nski, B., 1986, ApJ, 308, L43

\bibitem[\protect\astroncite{Paczy\'nski}{1990}]{pa90}
Paczy\'nski, B., 1990 ApJ, 348, 485

\bibitem[\protect\astroncite{Paczy\'nski}{1991}]{pa91}
Paczy\'nski, B., 1991, Acta Astr., 41, 257


\bibitem[\protect\astroncite{Peters}{1964}]{pet64}
Peters, P.~C., 1964, Phys. Rev., 136, 1224


\bibitem[\protect\astroncite{Phinney}{1991}]{phi91b}
Phinney, E.~S., 1991, ApJ, 380, L17

\bibitem[\protect\astroncite{Phinney \& Sigurdsson}{1991}]{ps91}
Phinney, E.~S., Sigurdsson, S., 1991, Nat, 349, 220


\bibitem[\protect\astroncite{Phinney \& Verbunt}{1991}]{pv91}
Phinney, E.~S., Verbunt, F., 1991, MNRAS, 248, 21

\bibitem[\protect\astroncite{Piran}{1996}]{pi96}
Piran, E.~S.,  1996, in Compact stars in binaries, 
ed. J. van Paradijs, 
E. P. J. van den Heuvel, E. Kuulkers (Dordrecht: Kluwer), p. 489

\bibitem[\protect\astroncite{Portegies~Zwart}{1995}]{pz95}
Portegies~Zwart, S., 1995, A\&A, 296, 691

\bibitem[\protect\astroncite{Portegies~Zwart \& Spreeuw}{1996}]{pzs96}
Portegies~Zwart, S.~F., Spreeuw, J.~N., 1996, A\&A, 312, 670

\bibitem[\protect\astroncite{Portegies~Zwart \& Verbunt}{1996}]{pzv96}
Portegies~Zwart, S.~F., Verbunt, F., 1996, A\&A, 309, 179

\bibitem[\protect\astroncite{Portegies~Zwart \etal}{1997}]{pzkr97}
Portegies~Zwart, S.~F., Kouwenhoven, M., Reynolds, A., 1997a, A\&A, 328, L2

\bibitem[\protect\astroncite{Portegies~Zwart \etal}{1997}]{pzve97}
Portegies~Zwart, S.~F., Verbunt, F., Ergma, E., 1997b, A\&A, 321, 207

\bibitem[\protect\astroncite{Portegies~Zwart \& Yungelson}{1997}]{pzy97}
Portegies~Zwart, S.~F., Yungelson, L. R., 1997, in preparation


\bibitem[\protect\astroncite{Rasio \& Livio}{1996}]{rl96}
Rasio, F., Livio, M., 1996, ApJ, 471, 366

\bibitem[\protect\astroncite{Ritossa \& Garc\'ia-Berro}{1996}]{rg96}
Ritossa, C., Garc\'ia-Berro., E., Iben, I. Jr., 1996, ApJ, 460, 489

\bibitem[\protect\astroncite{Sandage}{1986}]{sa86}
Sandage, A. 1986, A\&A, 161, 89

\bibitem[\protect\astroncite{Savonije}{1979}]{sav79}
Savonije, G.-J. 1979, A\&A, 71, 352

\bibitem[\protect\astroncite{Smarr}{1976}]{sb76}
Smarr, L.~L., Blandford, R., 1976, ApJ, 207, 574

  

  
\bibitem[\protect\astroncite{Stokes et~al.}{1985}]{std85}
Stokes, G., Taylor, J., Dewey, R., 1985, ApJ, 294, L21

\bibitem[\protect\astroncite{Tauris}{1996}]{tau96}
Tauris, T. M., 1996, A\&A, 315, 453

\bibitem[\protect\astroncite{Tauris \& Bailes}{1996}]{tb96}
Tauris, T. M., Bailes, M., 1996, A\&A, 315, 432  

\bibitem[\protect\astroncite{Tayler \etal}{1993}]{tml93}
Taylor, J. H., Manchester, R. N., Lyne, G., 1993, ApJS, 88, 529
 

\bibitem[\protect\astroncite{Thorne}{1996}]{th96}
Thorne, K. S., 1996, in  Compact stars in binaries, ed. 
  J. van Paradijs, E. P. J. van den Heuvel, E.
Kuulkers (Dordrecht: Kluwer), p. 153

\bibitem[]{}
Timmes, F.~X., Woosley, S.~E., Weaver, T.~A., 1996, 457, 834

\bibitem[\protect\astroncite{Tutukov \& Yungelson}{1973}]{ty73}
Tutukov, A. V.,  Yungelson, L. R., 1973, Nauchn. Informatsii, 27, 93 

\bibitem[\protect\astroncite{Tutukov \& Yungelson}{1978}]{ty78}
Tutukov, A. V.,  Yungelson, L. R., 1978, Nauchn. Informatsii, 41, 8

\bibitem[\protect\astroncite{Tutukov \& Yungelson}{1979}]{ty79}
Tutukov, A. V.,  Yungelson, L. R., 1979,
in Mass loss and evolution of O
stars, ed. C. De Loore, P. S. Conti (Dordrecht: Kluwer), p. 401

\bibitem[\protect\astroncite{Tutukov \& Yungelson}{1993a}]{ty923a}
Tutukov, A. V.,  Yungelson, L. R., 1993a, MNRAS, 260, 675 

\bibitem[\protect\astroncite{Tutukov \& Yungelson}{1993b}]{ty93b}
Tutukov, A. V.,  Yungelson, L. R., 1993b, SvA, 37, 411 

\bibitem[\protect\astroncite{Tutukov \etal}{1984}]{ty84}
Tutukov, A. V., Yungelson, L. R., 1994, MNRAS, 268, 871

\bibitem[\protect\astroncite{Tutukov \etal}{1984}]{tcy84}
Tutukov, A. V.,  Chugaj, N. N., Yungelson, L. R., 1984, SvALett, 
10, 244

\bibitem[\protect\astroncite{van~den Heuvel}{1976}]{vdh76}
van den Heuvel, E. P. J., 1976, in Structure and evolution of close binary 
stars, ed. P. Eggleton, S. Mitton, J. Whelan (Dordrecht: Kluwer), p. 35 

\bibitem[\protect\astroncite{van~den Heuvel}{1994}]{vdh94}
van~den Heuvel, E. P. J., 1994, A\&A, 1994, 291, L39 

\bibitem[\protect\astroncite{van~den Heuvel \& Heise}{1972}]{vdhh72}
van~den Heuvel, E. P. J., Heise, J., 1972, NatPhysSci, 239, 67

\bibitem[\protect\astroncite{van~den Heuvel \& Habets}{1984}]{hh84} 
van~den Heuvel, E.~P.~J., Habets, G., 1984, Nat, 309, 598

\bibitem[\protect\astroncite{van~den Heuvel \& Lorimer}{1996}]{vdhl96}
van~den Heuvel, E. P. J., Lorimer, D. R., 1996, MNRAS, 283, L37

\bibitem[\protect\astroncite{van den Heuvel \& van Paradijs} {1997}]{vdhvp97}
van den Heuvel, E.~P.J., van Paradijs J., 1997, ApJ, 483, 399

\bibitem[\protect\astroncite{van~den Heuvel \& kaper}{1994}]{vdhk94}
van~den Heuvel, E. P. J., Kaper, L., Ruymakers, E. 1994, in New
Horizon of X-ray Astronomy, eds. F. Makino, T. Ogashi (Tokyo, Iniversal
Acad. Press), p.\ 75

\bibitem[\protect\astroncite{van~den Hoek}{1997}]{vdh97}
van~den Hoek, B., 1997, PhD Thesis, U. Amsterdam 

\bibitem[\protect\astroncite{van den Hoek \& de Jong}{1997}]{vdkj97}
van den Hoek, B., de Jong T., 1997, A\&A, 318, 231

                                                                      

\bibitem[\protect\astroncite{Webbink}{1984}]{we84}
Webbink, R. F., 1984, ApJ, 277, 375 
 
\bibitem[\protect\astroncite{Wettig \& Brown}{1996}]{webr96}
Wettig, T., Brown, G. E., 1996, New Astronomy, 1, 17 

\bibitem[\protect\astroncite{White \& van Paradijs}{1996}]{wp96}
White, N. E., van Paradijs, J., 1996, ApJ, 473, L25

\bibitem[\protect\astroncite{Wielen}{1992}]{wiel96}
Wielen, R., 1992, in Landolt-B\"ornstein, Neue-Serie, b. 2, Astronomie 
und Astrophysik, p. 211  

\bibitem[\protect\astroncite{Wijers \etal}{1992}]{w92}
Wijers, R.~A. M. J., van Paradijs, J., van den Heuvel, E. P. J.,
1992, A\&A, 261, 145

\bibitem[\protect\astroncite{Wolszczan}{1990}]{wol90}
Wolszczan, A., 1990, IAU Circ.\ No.\ 5073
                                                        
\bibitem[\protect\astroncite{Woosley \etal}{1995}]{wlw95}
Woosley, S. E., Langer, N., Weaver, T. A., 1995, ApJ, 448, 315

\bibitem[\protect\astroncite{Yamaoka \etal}{1993}]{ysm93}
Yamaoka, H., Shigeyama, T., Nomoto, K., 1993, A\&A, 267, 433

\bibitem[\protect\astroncite{Yungelson \etal}{1995}]{y95}
Yungelson, L. R., Livio, M., Tutukov, A. V., Kenyon, S., 1995, ApJ, 447, 656

\bibitem[\protect\astroncite{Yungelson \etal}{1996}]{y96}
Yungelson, L. R., Livio, M.,  Tutukov, A. V., Truran, J. W.,  Fedorova, 
A. V., 1996, ApJ, 466, 890

\bibitem[\protect\astroncite{Zeldovich \etal}{1972}]{zin72}
Zel'dovich, Ya. B.,  Ivanova, N., Nadyozhin, D. K., 1972, SvA, 16, 209

\end{thebibliography}
\end{document}